\documentclass[aps,prd,showpacs,nofootinbib]{revtex4}
\usepackage{amssymb,graphics,graphicx, epsfig}
\usepackage{graphicx}
\usepackage{amsfonts}
\usepackage{amssymb}
\usepackage{slashed}
\usepackage{ulem}
\usepackage{bm}
\usepackage{color}
\usepackage{multirow}
\usepackage{array}
\usepackage{rotating}
\def\l[{\left[}
\def\r]{\right]}
\def\lp{\left(}
\def\rp{\right)}
\def\stau{\tilde{\tau}}

\newcommand{\beqn}{\begin{eqnarray}}
\newcommand{\eeqn}{\end{eqnarray}}
\newcommand{\be}{\begin{equation}}
\newcommand{\ee}{\end{equation}}
\newcommand{\non}{\nonumber\\}

\newcolumntype{x}[1]{
>{\centering}p{#1}}%
\newcommand{\tnhl}{\tabularnewline\hline}
\newcommand{\tn}{\tabularnewline}

\def\s1{$s_{\alpha}$}
\def\s2{$s_{\gamma}$}
\def\s3{$s_{\delta}$}
\def\c1{$c_{\alpha}$}
\def\c2{$c_{\gamma}$}
\def\c3{$c_{\delta}$}
\def\GeV{\text{ GeV}}

\newcommand{\wt}[1]{\widetilde{#1}}

\def \mh{m_{\frac{1}{2}}}
\newcommand{\gappeq}{\mathrel{\rlap {\raise.5ex\hbox{$>$}}
{\lower.5ex\hbox{$\sim$}}}}
\newcommand{\lappeq}{\mathrel{\rlap{\raise.5ex\hbox{$<$}}
{\lower.5ex\hbox{$\sim$}}}}

\newcommand{\TeV}      {~\mathrm{TeV}}
\begin{document}
\noindent
\title{ SUSY Discovery Potential and Benchmarks
for
  Early Runs at $\sqrt s =7$ TeV at the LHC}
\author{ {Baris~Altunkaynak,\footnote{altunkaynak.i@husky.neu.edu} Michael~Holmes,\footnote{holmes.mi@neu.edu} Pran Nath,\footnote{nath@neu.edu} Brent~D.~Nelson,\footnote{b.nelson@neu.edu} and Gregory Peim\footnote{peim.g@husky.neu.edu}
\vspace{.1cm}}
\vspace{.2cm}}
\affiliation{Department of Physics,
Northeastern University,  Boston, Massachusetts 02115, USA }
\begin{abstract}
We carry out an analysis of the potential of the {Large Hadron Collider (LHC)} to discover supersymmetry in runs  at $\sqrt s=7$ TeV with an accumulated luminosity of  (0.1--2)~fb$^{-1}$ of data.
The analysis is done with both {minimal supergravity} and  {supergravity)} models with nonuniversal soft breaking.  Benchmarks for
early discovery   with { (0.1--2)}~fb$^{-1}$ of data are given.
We provide an update of b-tagging efficiencies in PGS~4 appropriate for LHC analyses.
 A large number of signature
channels are analyzed  and it is shown that each
of the models exhibited are discoverable at the 5$\sigma$ level or more above the standard model
background in {\it several}  signature channels which would provide cross checks for a  discovery of supersymmetry.
It is shown that some of the benchmarks are discoverable with 0.1~fb$^{-1}$ of data again
with detectable signals in several channels.
\end{abstract}

\maketitle

 \preprint{YITP-SB-09-23}
\section{Introduction\label{intro}}
{The Large Hadron Collider (LHC)} is currently running and collecting data at $\sqrt s=7$ TeV. It is expected  that it would continue
 until  it has collected {at least} 1~fb$^{-1}$ of integrated luminosity.
It is then interesting to investigate the discovery potential of LHC for supersymmetry (SUSY)  in these
early runs.\footnote{For a review of the discovery potential at $\sqrt s=14$ TeV and $\sqrt s =10$ TeV see~\cite{Nath:2010zj} and the references therein.}
Some work in this direction has already been done~{\cite{Bauer:2009cc,Baer:2010tk,Bhattacharyya:2010gm,Alves:2010za,Desch:2010gi,Edsjo:2009rr}}.  
 We carry out the analysis
of the SUSY discovery potential  of the LHC
using the {next-to-lightest supersymmetric partner (NLSP)} as a discriminant of models.
{Several possibilities arise, the most dominant of which involve a
$\chi^{\pm}$, $\tilde \tau$, $\tilde t$, CP odd Higgs $A$ or $\tilde g$ as the NLSP}.
 As is well known, the parameter space of minimal supersymmetric standard model
 (MSSM) is very large, consisting of more than a hundred
 parameters, and thus an analysis within this full space of MSSM is intractable. This space is
 significantly reduced in well motivated models such as the minimal supergravity model (mSUGRA) where
 one assumes R parity invariance and
  defines the symmetry breaking parameters at the grand unification (GUT) scale. These are evolved
  to low  scales by renormalization group evolution
   where the entire sparticle spectrum is  determined  in terms of the GUT scale parameters.
 Specifically in mSUGRA under the constraint of R parity one has four parameters and one sign [see Eq.(3)] and in nonuniversal
 supergravity models with nonuniversalities in the gaugino sector the space is extended to six parameters
 and a sign.  The allowed parameter space is subjected to further theoretical and experimental
 constraints, i.e., constraints of radiative electroweak symmetry breaking (REWSB) with color and charge
 conservation,
 satisfaction of flavor
 changing neutral current
 (FCNC) constraints, and experimental lower limit constraints on the  Higgs masses and on the masses of the
 sparticles. Further, in supergravity unified models under the constraints of R parity invariance
 the lowest R parity odd particle (LSP) turns out to be the neutralino over a large part of the parameter space
 and thus a possible candidate for dark matter. Imposition of the WMAP constraint and constraints from
 direct detection of dark matter further reduce the allowed parameter space in these supergravity models.

However, even after the imposition of all the constraints the allowed parameter space is still large
and additional criteria must be used in model selection  for  a study of their
signatures.  This  study is focussed on SUSY discovery in the early runs at the LHC, i.e., at
$\sqrt s=2$ TeV with (1-2) fb$^{-1}$ of integrated luminosity.
Obviously then it makes sense to choose those SUSY models which  can be explored
within the parameters of the data that will be collected in these early runs. However, within this
general rubric one ought to make the search as large as possible. For example, there is no theoretical
necessity to limit our analysis exclusively to just one region of REWSB such as just to the
 stau coannihilation branch. Thus in our analysis, under the constraint that the model points we explore
 be observable in the early runs, we search for a diverse set of models where the NLSP could be
 one of the allowed set by REWSB.  For the analysis we present here we find this set to be
 a chargino ($\chi^{\pm}$), a stau ($\tilde \tau$), a gluino ($\tilde g$), CP odd Higgs (A$^0$), and a stop
 ($\tilde t$). Thus in the analysis of this paper we give several benchmarks with this diverse set of
 NSLPs.   It should be kept in mind  that there is still a significant amount of subjectivity in the choice
 of the benchmarks. To mitigate the subjective element in the choice of the benchmarks we further choose the input
 parameters
 to be as diverse as possible. 
 The investigation is
done both for mSUGRA  as well as for models with nonuniversalities.
The outline of the rest of the paper is as follows: In Sec.~(\ref{background}) we discuss the analysis of standard model
backgrounds at $\sqrt s =7$ TeV.  In Sec.~(\ref{sparticle}) we
give an analysis of the cross sections for  the production of various sparticle processes. These include
the production of  gluinos and squarks,
 production of  a combination of  gluinos and charginos or neutralinos,  and production of charginos and neutralinos.
  In Sec.~(\ref{signatureAnalysis}) we give an analysis of possible signatures,
 discuss the SUSY discovery potential and give reach plots of the LHC early runs.
  In { Sec.~(\ref{modelSim})} we  give several benchmarks  consistent with all current constraints which
 are possible candidates for early discovery.   Conclusions are given in Sec.~(\ref{conclusion}) {and larger tables have been relocated to Sec.~(\ref{tables})}.

 \section{Analysis of standard model Backgrounds at $\sqrt s = 7$ TeV\label{background}}
 A central element in the discovery of new physics is the determination of the standard model {(SM)} backgrounds
for the processes where one expects to see  new physics. One such analysis at $\sqrt s=7$ TeV
has already appeared in the literature~\cite{Baer:2010tk}.\footnote{For some previous works on early
discovery though at higher energies see{~\cite{Hubisz:2008gg,Dreiner:2010gv,Baer:2008ey,Dietrich:2010tf, Bhattacharyya:2010gm}}}
Here we  give an independent analysis of the relevant
backgrounds.
In our analysis, we use MadGraph~4.4~\cite{Alwall:2007st} for  parton level processes, { Pythia{~6.4}~\cite{pythia} for  hadronization, and PGS~4 for detector simulation~\cite{pgs}.  We used MLM matching with a $k_T$ jet clustering scheme to prevent double counting{ of final states}{.   Furthermore, the $b$-tagging efficiency of} PGS~4 was updated to {better represent}  the  analysis at the LHC.  A discussion of this improvement is given {below}.
 The result of our analysis is  presented in Table~(\ref{smprocesses}) with parton level
 cuts as specified in Eq.~(\ref{cuts}) and in
 the caption of Table~(\ref{smprocesses}).  Our analysis compares well with the analysis of~\cite{Baer:2010tk}.
 The differences between the two analyses are in part due to the
 differences between matrix-element Monte Carlo generators which are known to exist~\cite{Alwall:2007fs}.  For example, we have exclusively used MadGraph with a $k_T$-based jet clustering algorithm in our analysis, while AlpGen~\cite{Mangano:2002ea} uses a cone-based jet clustering algorithm and was the dominant tool used in~\cite{Baer:2010tk}.
 }
In the analysis here we have used  CTEQ6L1~\cite{Pumplin:2002vw}
  parton distribution functions  for generating the SM background, and a basic cut was applied
  such
  that all final state partons (except the  top quarks) are required to have $p_{T}>40\text{ GeV}$.

\begin{table}[t!]
\centering
{\bf\boldmath  A display of the processes analyzed and their
 standard model backgrounds at $\sqrt s=7$ TeV}
 \vspace{.3cm}

\begin{tabular}{|l||c|c|c|}
\hline
  \multirow{2}{*}{SM process} & Cross  & Number & Luminosity  \\
&section (fb)&of events & $\left(\text{fb}^{-1}\right)$\\\hline
$\text{QCD } 2, 3, 4 \text{ jets (Cuts1)}$ &$2.0\times 10^{10}$  & $74 \text{M}$ &$3.7\times10^{-3}$\\
$\text{QCD } 2, 3, 4 \text{ jets (Cuts2)}$&   $7.0\times 10^{8}$& $98 \text{M}$ &$0.14$ \\
$\text{QCD } 2, 3, 4 \text{ jets (Cuts3)}$& $4.6 \times 10^{7}$ & $40\text{M}$&$0.88$ \\
$\text{QCD } 2, 3, 4 \text{ jets (Cuts4)}$& $3.9\times 10^{5}$& $1.7\text{M}$&$4.4$\\
 $t\bar{t}+0,1,2\text{ jets}$ &  $1.6\times 10^{5}$ & $4.8\text{M}$& $30$\\
 $b\bar{b}+0,1,2\text{ jets}$ &  $9.5\times 10^{7}$ & $95\text{M}$&$1.0$\\
 $Z/\gamma\left(\to \ell \bar{\ell}, \nu \bar{\nu}\right)+0,1,2,3\text{ jets}$ &  $6.2\times 10^6$ &$6.2\text{M}$&$1.0$\\
 $W^{\pm}\left(\to \ell\nu\right)+0,1,2,3\text{ jets}$ &  $1.9\times 10^7$ &$21\text{M}$&$1.1$\\
  $Z/\gamma\left(\to \ell \bar{\ell}, \nu \bar{\nu}\right)+t\bar{t}+0,1,2\text{ jets}$ &  $56$ &$1.0\text{M}$&$1.7\times10^{4}$\\
  $Z/\gamma\left(\to \ell \bar{\ell}, \nu \bar{\nu}\right)+b\bar{b}+0,1,2\text{ jets}$ &  $2.8\times 10^3$ &$0.1\text{M}$&$36$\\
 $W^{\pm}\left(\to \ell\nu\right)+b\bar{b}+0,1,2\text{ jets}$ &  $3.2\times 10^3$ &$0.6\text{M}$&$1.8\times 10^2$\\
 $W^{\pm}\left(\to \ell\nu\right)+t\bar{t}+0,1,2\text{ jets}$ &$70$ &$4.6\text{M}$&$6.5\times 10^{4}$\\
  $W^{\pm}\left(\to \ell\nu\right)+t\bar{b}\left(\bar{t}b\right)+0,1,2\text{ jets}$ & $2.4\times10^2$ &$2.1\text{M}$&$8.7\times10^{3}$\\
$t\bar{t}t\bar{t}$& $0.5$ & $0.09\text{M}$&$1.8\times10^{5}$\\
$t\bar{t}b\bar{b}$& $1.2\times 10^2$ & $0.32\text{M}$ &$2.7\times10^3$\\
$b\bar{b}b\bar{b}$ & $2.2\times 10^{4}$ & $0.22\text{M}$&$1.0$ \\
$W^{\pm}\left(\to\ell \nu\right)+W^{\pm}\left(\to\ell \nu\right)$ & $2.0\times 10^3$&$0.05\text{M}$&$25$\\
$W^{\pm}\left(\to\ell \nu\right)+Z\left(\to~all\right)$ & $1.1\times 10^3$&$1.3\text{M}$&$1.1\times10^{3}$\\
$Z\left(\to~all\right)+Z\left(\to~all\right)$ & $7.3\times10^{2}$&$2.6\text{M}$&$3.6\times10^3$\\
$\gamma + 1, 2, 3 \text{ jets}$ & $1.5\times 10^{7}$ & $16\text{M}$ & $1.1$ \\\hline
\end{tabular}
\caption{An exhibition of the standard model  backgrounds  computed in this work at $\sqrt{s}=7\text{ TeV}$.  {All processes were generated using MadGraph~{4.4}~\cite{Alwall:2007st}.  Our notation here is that $\ell=e,\mu,\tau$, and $all=\ell,\nu, jets$.  Cuts1-Cuts4 indicated in the table
are defined in Eq.~(\ref{cuts}).
In the background analysis we eliminate
 double counting between the process $W^{\pm}+t\bar{b}\left(\bar{t}b\right)$ and $t\bar{t}$ by subtracting out double resonant diagrams of $t\bar{t}$ when calculating $W^{\pm}+t\bar{b}\left(\bar{t}b\right)$.$^{3}$
}}
\label{smprocesses}
\end{table}
\footnotetext[3]{{When studying $W+\bar{t}b\left(t\bar{b}\right)$ processes there is a potential to  double count such final states if one also considers $t$, $\bar{t}$  prodcution processes.  To prevent this  double counting we have eliminated all diagrams involving a top quark from the set of diagrams that lead to $W+\bar{t}b$ final states, with an analogous requirement for $W+t\bar{b}$ production.}}

\beqn
\text{Cuts1}&=&40\text{ GeV} <E_{T}\lp j_1\rp < 100\text{ GeV},
~~\text{Cuts2} =100\text{ GeV} <E_{T}\lp j_1\rp < 200\text{ GeV},   ~~~{\rm Parton ~level}  \non
\vspace{0.4cm}
\text{Cuts3} &=& 200\text{ GeV} <E_{T}\lp j_1\rp < 500\text{ GeV},
~~\text{Cuts4} =  500\text{ GeV} <E_{T}\lp j_1\rp < 3000\text{ GeV} ~~~~~{\rm cuts}.
\label{cuts}
\eeqn

An important {object in many possible SUSY discovery channels is a} b-tagged jet.
    In PGS~4   $b$ tagging
     is done  based off of the Tevatron $b$-tagging efficiency.
 However, it is pertinent to ask if this is a valid approximation of what is to
  be expected at the LHC.  Using the Technical Design Reports  (TDR) of CMS~  \cite{Bayatian:2006zz} and of ATLAS~\cite{Aad:2009wy} one can extract  the
  expected {b-tagging} efficiency as a function of {jet pseudorapidity}, $\eta$, and transverse energy  $E_T${ for jets originating from heavy-flavor partons}.
   In  Fig.~(\ref{pgsbtag}) we give a comparison of the functions given in the ATLAS  TDR~\cite{Bayatian:2006zz,Aad:2009wy}
     with the one given in PGS~4.  {We have omitted the CMS data, since in the TDR {b tagging is binned into two sets
     of
      $\eta$ so one could not extract a continuous function from it.}}
     The left panel of Fig.~(\ref{pgsbtag}) gives  the b-tagging efficiencies as a function of $E_T$
     for ATLAS and the so called ``tight'' and ``loose'' efficiencies  as defined in  PGS~4.
     One finds
     a significant difference between these  and those expected in the
     ATLAS and CMS detectors.  A similar analysis but as a function of $\eta$ is given in the right
     panel of Fig.~(\ref{pgsbtag}) where the difference between the efficiencies given by PGS~4 and
     by {(normalized)} ATLAS is even more glaring. {For this analysis, we do not extend the definition of b-tagging
     beyond
      $\eta$ of $2.0$.  {Previ ously, b-tagging efficiencies were assumed to approach a constant value for $E_T \geq 160 \GeV$.  We have extended this such that the tagging efficiency reaches a constant for $E_T \geq 300 \GeV$. }  Thus we have
  updated $b$-tagging functions as given in  Eq.~(\ref{btageqn}) where we have kept the same degree polynomial as  the {preexisting} {$b$-tagging functions in PGS~4.  {Furthermore, the total b-tagging efficiency is the product of the $E_T$ and $\eta$ functions listed in Eq.~(\ref{btageqn}).  Here we make no modification to the default PGS~4 rate for mistagging b jets.}  Our revised
 b-tagging functions have the form
 \begin{eqnarray}
\nonumber b_{E_T}&=&0.0781391+0.0202661E_T-0.000259664E_T^2+1.5509\times10^{-6}E_T^3-4.46698\times 10^{-9} E_T^4 + 4.7995 \times 10^{-12}E_T^5 \\
\vspace{0.4cm}
b_{\eta} &=&  1.00885-0.0497485\eta +0.693036\eta^2-0.0361142\eta^3-0.0222204\eta^4+0.00797621\eta^5. \label{btageqn}
\end{eqnarray}
\begin{figure}[t!]
{\bf\boldmath A comparison of $b$-tagging efficiency in PGS~4 vs ATLAS detectors}
\vspace{-1.3cm}
\begin{center}
\includegraphics[scale=0.25]{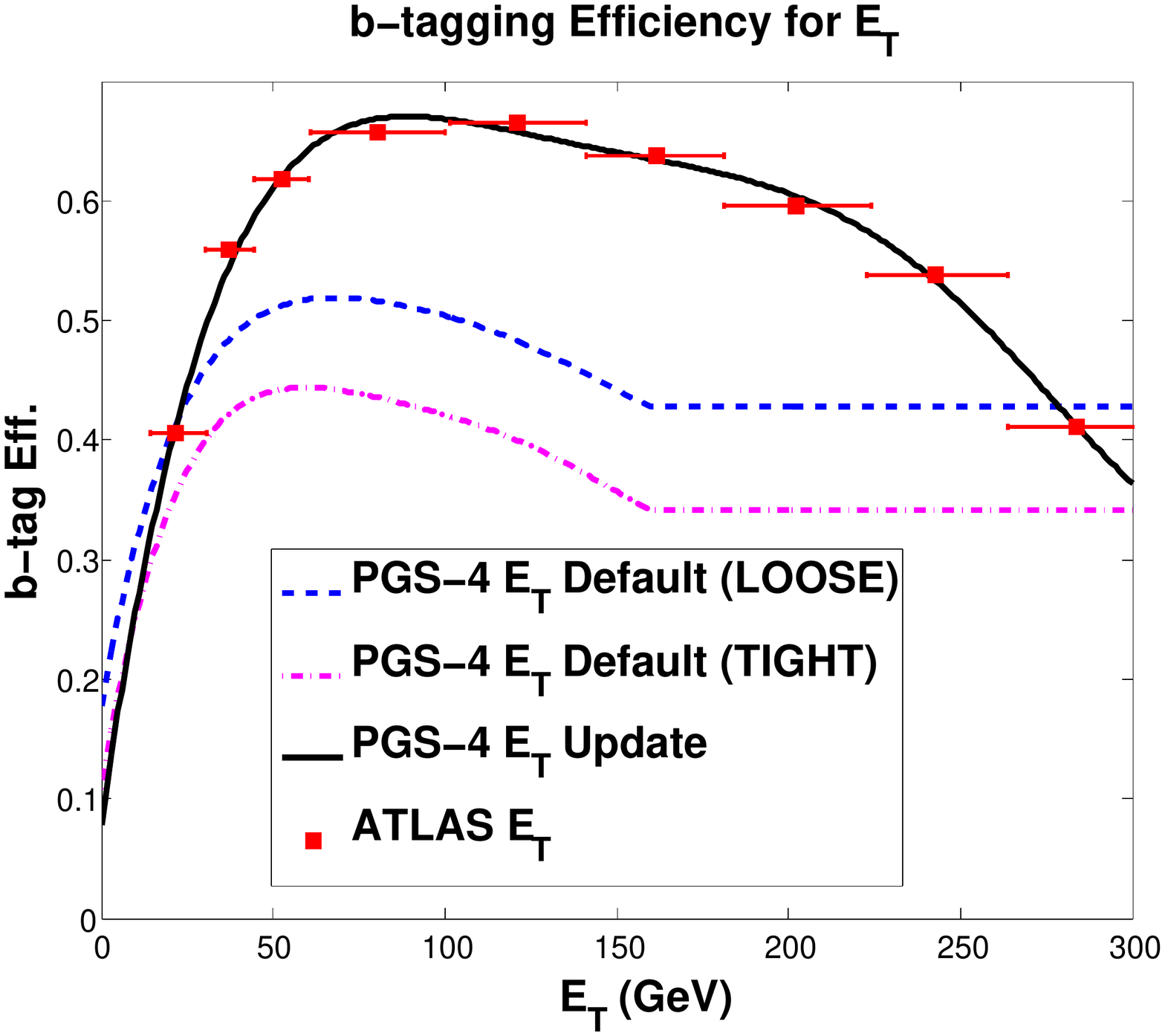}
\hspace{1cm}
\includegraphics[scale=0.25]{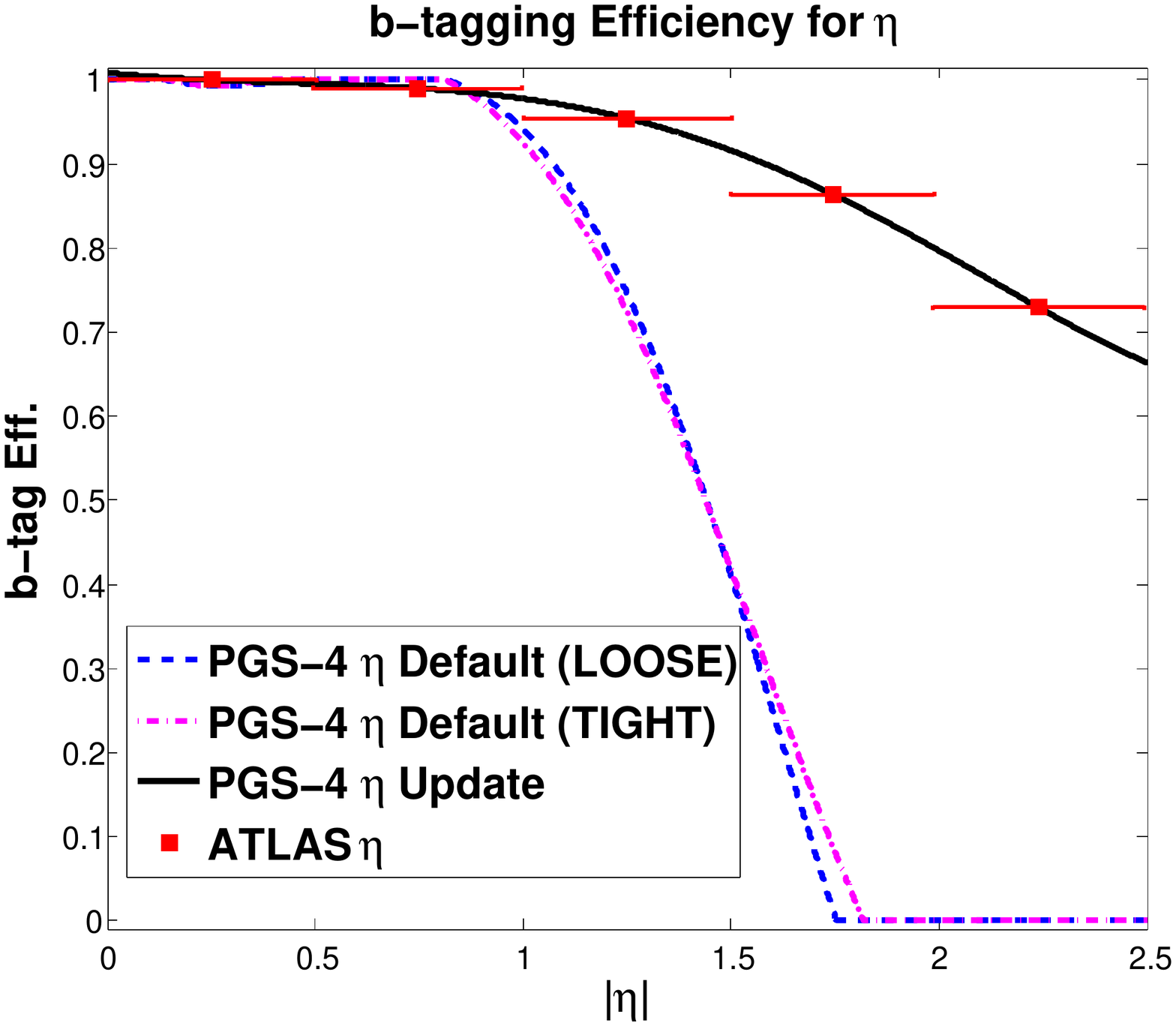}
\vspace{-1.5cm}
\caption{ Left panel: A comparison of the $b$-tagging efficiency of ATLAS  and the
loose and tight efficiencies of PGS~4 as a function of $E_T$. Right panel:
A comparison of the $b$-tagging efficiency of ATLAS  and the
loose and tight efficiencies of PGS~4 as a function of $\eta$. Ours fits to the efficiency of ATLAS as a function of $E_T$,
and $\eta$ as parametrized by Eq.~(\ref{btageqn}) are also exhibited.}
\label{pgsbtag}
\end{center}
\end{figure}

\noindent
\section{Sparticle production cross sections at $\sqrt s =7$ TeV \label{sparticle}}

The soft breaking sector of the MSSM is rather large consisting of over a hundred arbitrary parameters.
Here we use the framework of high scale supergravity to reduce this arbitrariness.
Thus, for the supergravity grand unified model with universal soft breaking, i.e.,
  mSUGRA~\cite{msugra},  one has  just four parameters and the sign
of the Higgs mixing parameter $\mu$; i.e., one  has
\beqn
m_0, m_{\frac{1}{2}}, A_0, \tan\beta, {\rm sign}(\mu),
\label{msugra}
\eeqn
where $m_0$ is the universal scalar mass, $m_{\frac{1}{2}}$ is the universal gaugino mass,
$A_0$ is the coefficient of the trilinear coupling, and $\tan\beta$ is the ratio of two Higgs vacuum expectation values in {the} MSSM. Since the nature of physics at the Planck scale is still largely unknown, one may extend
 the minimal supergravity model to include nonuniversalities in the gaugino and
 Higgs sectors as well as in the flavor  sector consistent with the FCNC constraints.  One of the
 most widely used extensions of mSUGRA consists of the inclusion of nonuniversalities in the gaugino
 sector~\cite{NUSUGRA,WinoSUGRA,NUSUGRAColliders,Choi:2007ka,Altunkaynak:2009tg,Feldman:2010uv,Feldman:2010wy}.
 Thus one may include these nonuniversalities  by parametrizing
 the gaugino masses  at the  grand unified scale, {which we take to be $10^{16} \GeV$, via the relations} $\tilde m_i =m_{1/2}(1+ \delta_i)$ (i=1,2,3) corresponding
 to the gauge groups $U(1), SU(2)_L, and SU(3)_C$.

  {In this framework, the} model points are generated by the imposition of REWSB, particle mass limits from
 LEP and the Tevatron, relic density constraints from WMAP~\cite{Jarosik:2010iu},  the $g_{\mu}-2$ constraints, and
  FCNC constraints  {from} $B_s\to \mu^+\mu^-$ and
  $b\to s+\gamma$.
    WMAP has measured $\Omega_{DM} h^2$ to a great accuracy{ with} $\Omega_{DM}h^2=0.1109\pm 0.0056$~\cite{Jarosik:2010iu}.  However,
   to account for the errors in the theoretical computations and possible variations in the computation of the relic density using different codes we take a rather
    wide range in the relic density  constraints, i.e.,
    $0.06<\Omega_{DM}h^2<0.16$, in our analysis.
   Regarding the $g_{\mu}-2$ constraint~\cite{Bennett:2004pv}  recent analysis of  the hadronic corrections~\cite{Davier:2009zi}
indicate a significant deviation   around $3.9\sigma$ between the SM prediction and experiment. Such a contribution can
arise from supersymmetry~\cite{yuan}   and the size of the correction indicates the {supersymmetric particles (sparticles)} to be low{ in mass}.

\begin{figure}[!t]
\begin{center}
{\bf\boldmath Sparticle Production Cross Sections of mSUGRA at $\sqrt s=7$ TeV}\\
\vspace{.3cm}
\includegraphics[scale=0.478]{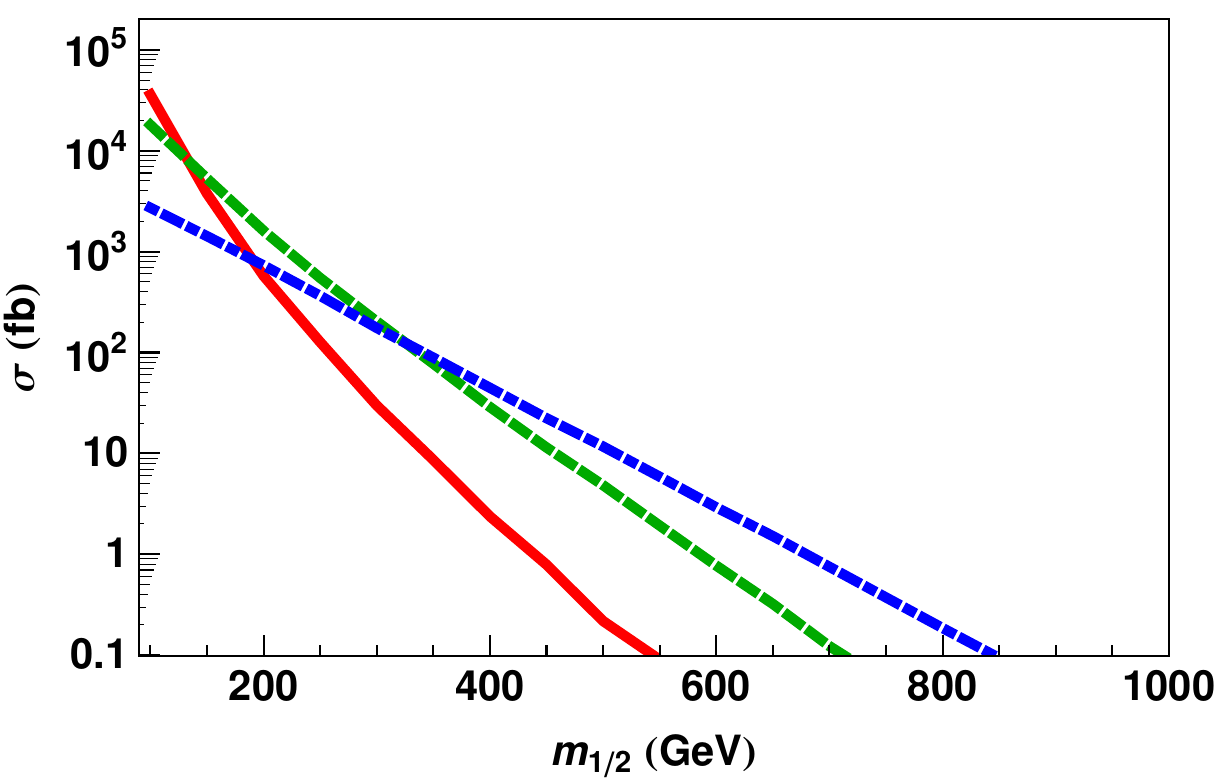}
\includegraphics[scale=0.478]{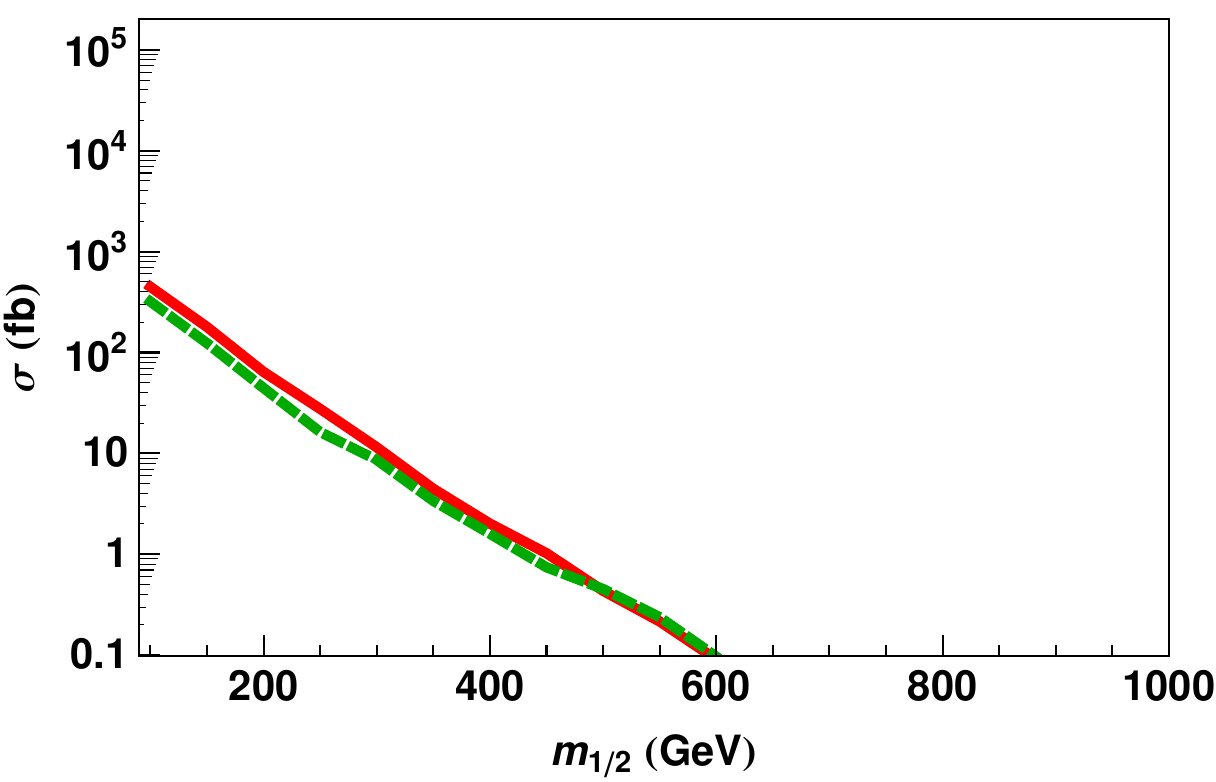}
\includegraphics[scale=0.478]{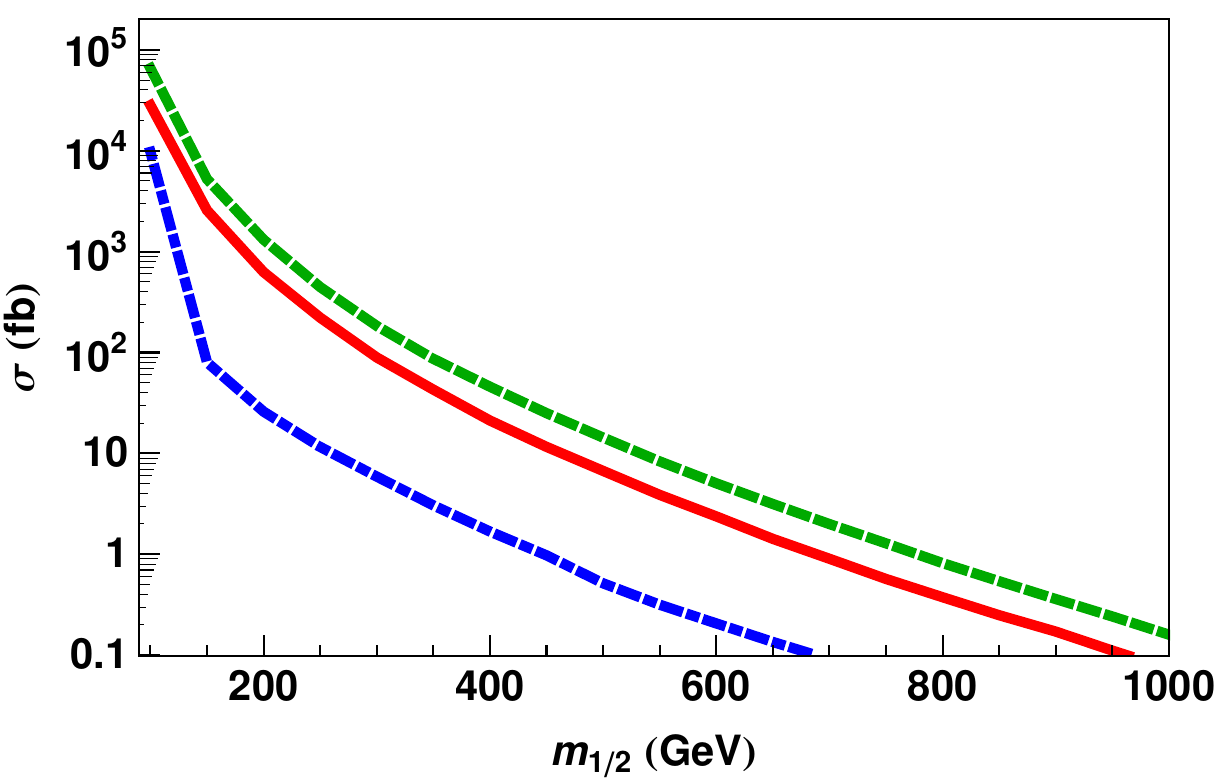}
\caption{An exhibition of the sparticle  production cross sections at the LHC at $\sqrt s=7$ TeV
for mSUGRA  as a function of the universal gaugino mass $m_{1/2}$   at the GUT scale
when $m_0=500$ GeV, $A_0=0$, $\tan\beta=20$ and sign($\mu$)= +1.
Left panel: production cross sections of $\tilde{g}\tilde{g}$, $\tilde{g}\tilde{q}$, $\tilde{q}\tilde{q}$ (solid red, dashed green, dashed blue lines). Middle panel: production cross sections for $\tilde{g}{\chi}^\pm$, $\tilde{g}{\chi}^0$  (solid red, dashed green lines).
Right panel: production cross sections for  ${\chi}^\pm{\chi}^\pm$, ${\chi}^\pm{\chi}^0$, ${\chi}^0{\chi}^0$ (solid red, dashed green, dashed blue lines).\label{fig:1}}
\end{center}

\end{figure}

\begin{figure}[!h]

{\bf \boldmath Contours of  $\sigma_{SUSY}$ in the $m_{\tilde g}-m_{\chi^{\pm}}$ mass plane at $\sqrt s=7$ TeV}\\
\vspace{.3cm}
\includegraphics[scale=0.505]{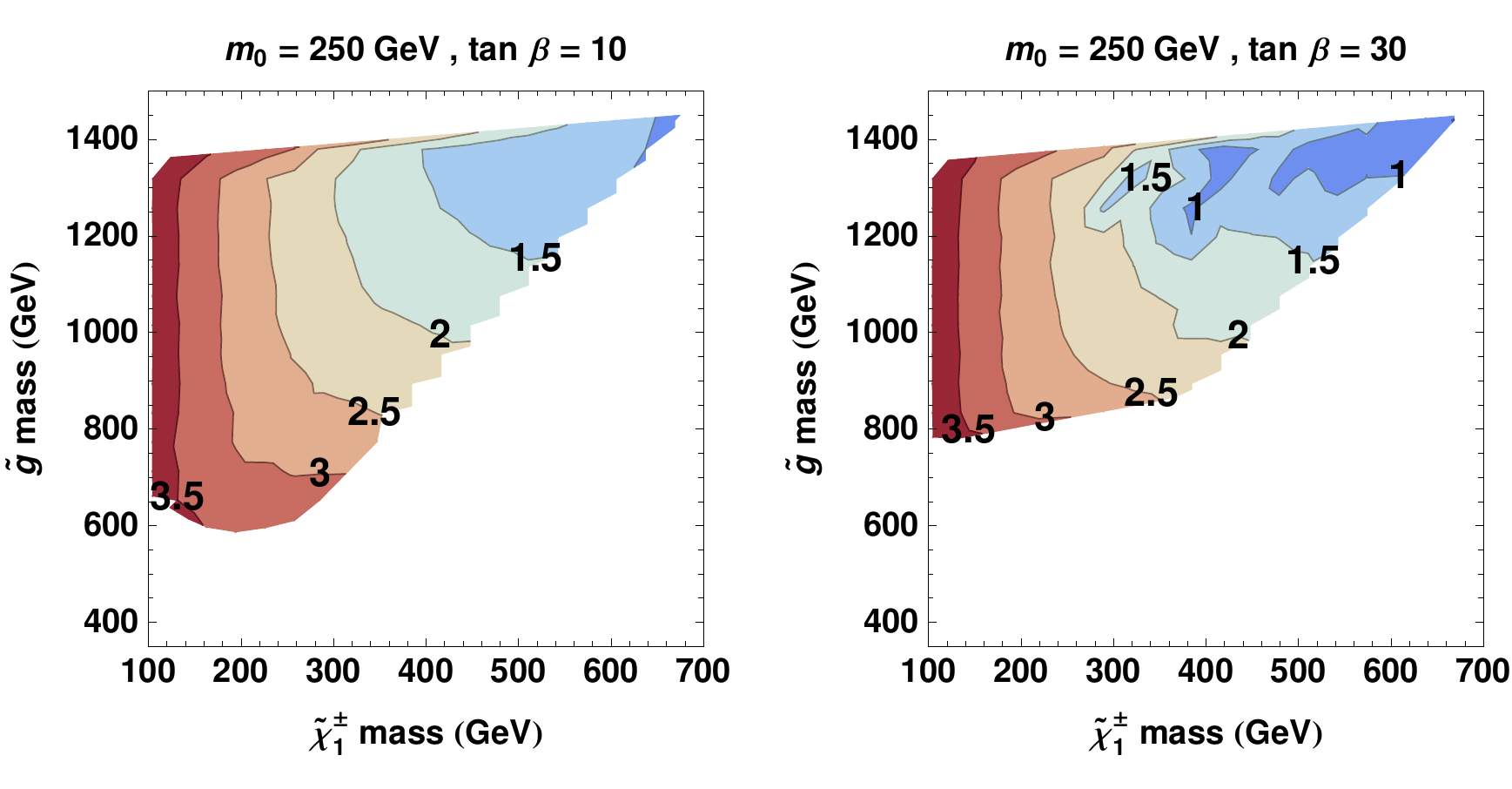}
\includegraphics[scale=0.505]{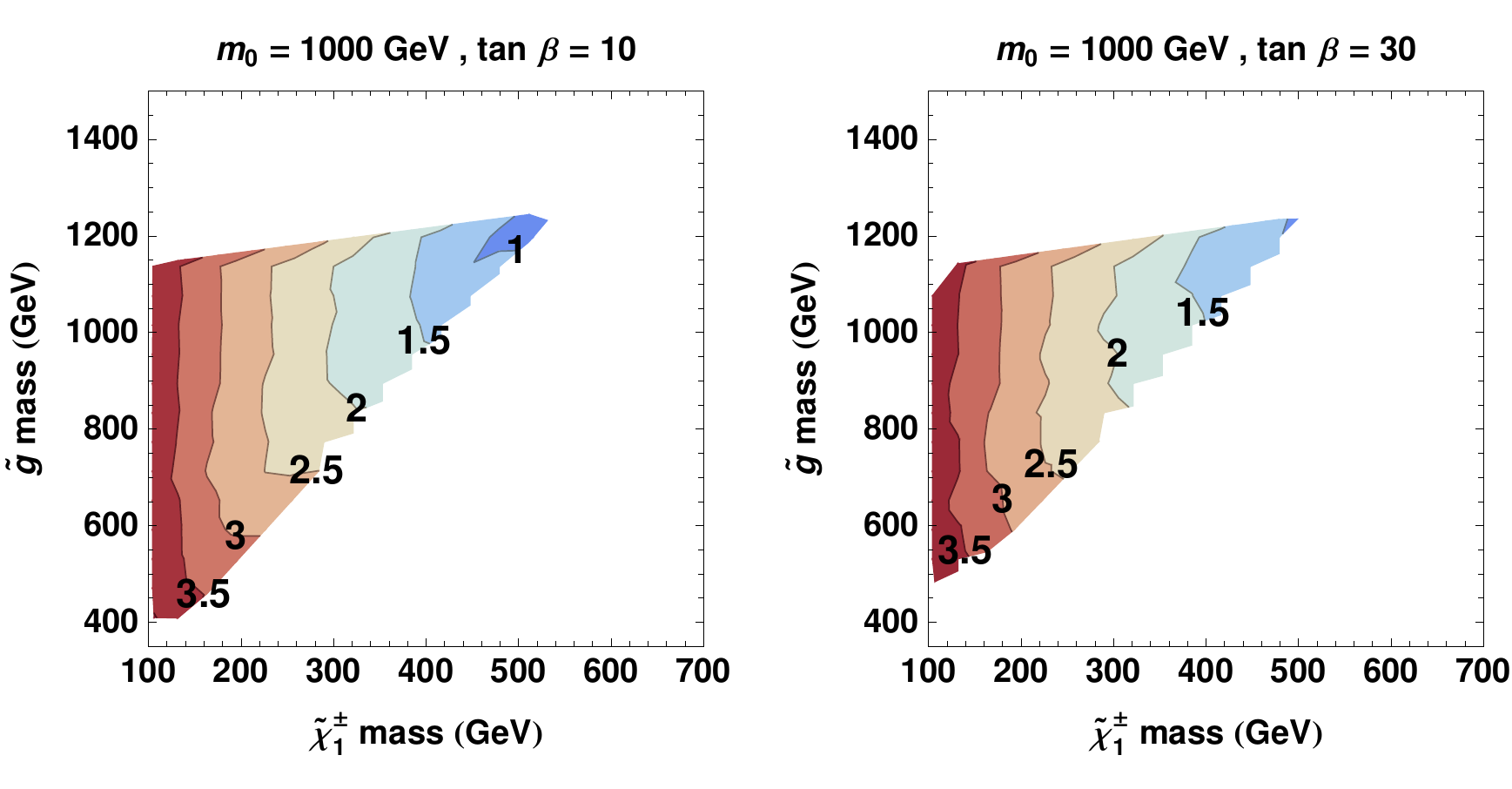}
\caption{Contour plots with constant values of  ${\log}(\sigma_{SUSY}/\text{fb})$
for $\sigma_{SUSY}$
in $m_{\tilde g}-m_{\chi^{\pm}}$ mass plane for the
case with nonuniversalities in the gaugino sector. {Gaugino masses $m_1$, $m_2$, and $m_3$ vary up to 1 TeV.}
First panel from left: $m_0 = 250 \GeV$, $\tan\beta = 10$ while $A_0 = 0$, sign($\mu$)=+1;
second panel from left: $m_0 = 250 \GeV$, $\tan\beta = 30$;
third panel from left: $m_0 = 1000 \GeV$, $\tan\beta = 10$;
fourth panel from left: $m_0 = 1000 \GeV$, $\tan\beta = 30$. \label{contour}}
\end{figure}
\vspace{.1cm}

For the FCNC process $B_s\to \mu^+\mu^-$ we take the constraint to be
 $\mathcal{BR}\left(B_s\to \mu^+\mu^-\right)<5.8\times 10^{-8}$~\cite{:2007kv,Amsler:2008zzb} and for the
 process $b\to s\gamma$ we take the constraint to be
     $ \mathcal{BR}\left(b\to s\gamma\right)=\left(352\pm 34\right)\times 10^{-6}$~\cite{Barberio:2008fa, Misiak:2006zs}.
     We note in passing that  currently there is a small discrepancy between the SM prediction and the
     experimental result for $b\to s \gamma$ which is a possible hint for the
     SUSY contribution~\cite{susybsgamma}  and
     hence another indication of possible low-lying sparticle masses. Thus this discrepancy along with the reported
     $g_{\mu}-2$ result
         is encouraging for an early SUSY discovery~\cite{Chen:2009cw}.
            In addition to the above, LEP and Tevatron mass
         constraints on the sparticle masses and on the Higgs masses are applied. These are
       { $m_A >85\text{ GeV}$, $m_{H^{\pm}}>79.3\text{ GeV}$, $m_{\tilde{t}_1}>101.5\text{ GeV}$, and $m_{\stau_1}>98.8\text{ GeV}$ where $A$ is the CP odd Higgs and $H^{\pm}$ is the charged Higgs.  Further, we
impose the lightest CP even Higgs mass
constraint~\cite{Djouadi:2001yk} $m_h>~\left(93.5
+15x+54.3x^2-48.4x^3-25.7x^4+24.8x^5-0.5\right)\text{ GeV}$ where
$x=\sin^2\left(\beta-\alpha\right)$ and $\alpha$ is the Higgs mixing
angle. {The final term in the bound represents a
theoretical error of $0.5 \GeV$ in the calculation of $M_h$ and
$M_A$ assumed by the authors of Ref.~\cite{Djouadi:2001yk}.}
Additionally we use the constraints  $m_{\chi_{1}^{\pm}}>104.5\text{
GeV}$ if $\left|m_{\chi_1^{\pm}}-m_{\chi_1^o}\right|> 3\text{ GeV}$
for the chargino mass and $m_{\tilde{g}}>309\text{ GeV}$  for the
gluino mass ~\cite{Amsler:2008zzb}. In the analysis we use a top
(pole) mass of $m_t=173.1$ GeV. Finally, {we also apply the
constraint that the model points are consistent with recent data
from CDMS-II~\cite{Ahmed:2008eu} and XENON-100~\cite{Aprile:2010um}.
Using MicrOMEGAs{~2.4}~\cite{Belanger:2008sj} the spin independent
neutralino-proton
             cross section was calculated and compared to CDMS-II and XENON-100. { Furthermore, we compare our results to the expected sensitivity { for XENON-100 6000 ~kg $\times$ day and XENON-1Ton for 1 ton $\times$ year~\cite{futureXENON}  as well as the expected sensitivity for SuperCDMS~\cite{futureSCDMS}.}}

Next  we present the  cross sections for  sparticle production  processes in mSUGRA  {calculated from} Pythia{~6.4}  at $\sqrt s=7$ TeV.   This is done by generating  5K events for multiple $m_{1/2}$ values  where the other parameters  are taken to be  $m_0 = 500 \GeV$,     $A_0=0$, $\tan \beta = 20$, and  $\mu>0$.   The results are exhibited in Fig.~(\ref{fig:1}) where the left panel  gives the  cross sections for the production   of $\tilde{g}\tilde{g}$ (solid red line), $\tilde{g}\tilde{q}$ (dashed green line), $\tilde{q}\tilde{q}$(dashed blue line) as a  function of $m_{1/2}$.
The middle panel gives the cross sections for  the production of   $\tilde{g}{\chi}^\pm$ (solid red line), $\tilde{g}{\chi}^0$ (dashed green line), and the right panel gives the  production  cross section for ${\chi}^\pm{\chi}^\pm$ (solid red line), ${\chi}^\pm{\chi}^0$ (dashed green line), ${\chi}^0{\chi}^0$ (dashed blue line). The analysis of Fig.~(\ref{fig:1})  shows these cross sections to be significant, indicating that at low mass scales as many as $10^4$ or more SUSY events will be generated with 1~fb$^{-1}$ of integrated luminosity at the LHC.
 A  similar analysis for the case with  nonuniversalities in the gaugino mass sector is given in Fig.~(\ref{contour}), where
 we give contour plots  in the $m_{\tilde g}-m_{\chi^{\pm}}$} mass plane with other parameters
as stated in the caption of the figure. The plots  give contours of constant {$\log\left(\sigma_{SUSY}/\text{fb}\right)$} in the range
$1-3.5$. The contours' plots indicate that a chargino mass up to
about 500 GeV and a gluino mass up to roughly  1 TeV would give up to $10^{3}$ or more
 events with 1~fb$^{-1}$ of integrated luminosity.  Of course,  the discovery of  sparticles  requires
 an analysis including the backgrounds and selection of appropriate cuts to enhance the signal to the
 background  ratio. This will be discussed in Secs.~(\ref{signatureAnalysis}) and (\ref{modelSim}).

 \section{Signature Analysis and SUSY Discovery Reach  at $\sqrt s=7$ TeV and 1~fb$^{-1}$\label{signatureAnalysis}}

\begin{figure}[!t]
\begin{center}
{\bf\boldmath Reach Plot at $\sqrt s=7$ TeV up to 2~fb$^{-1}$ of Integrated Luminosity.\vspace{0.3cm}}
\includegraphics[scale=0.75]{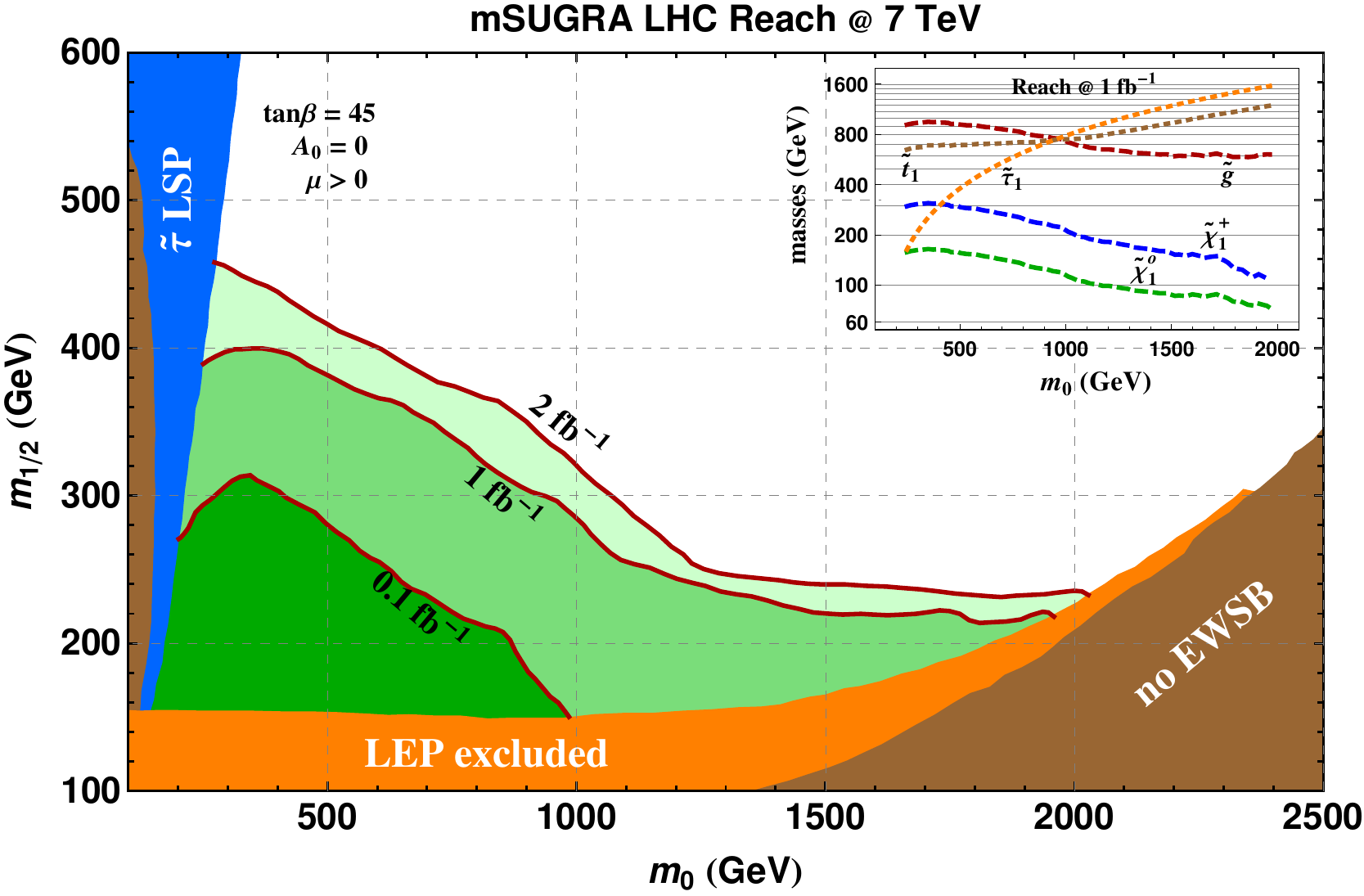}
\caption{ A reach plot for mSUGRA
using the signature analysis and the standard model backgrounds  of this work  given in
Table~(\ref{smprocesses}) at the LHC with $\sqrt s=7$ TeV and  1~fb$^{-1}$ of integrated luminosity.
The mSUGRA parameters used are $A_0=0$, $\tan\beta =45$, sign$(\mu)$ =1.
The analysis is done under the conditions of REWSB and the LEP and Tevatron constraints but
without the imposition of the  relic density and FCNC constraints.
The condition used for a signal to be observable is
 $S > \text{max}(5\sqrt {SM},10)$  where $SM$ stands for the standard model background.
 Early LHC  reaches at 1fb$^{-1}$ for the
  gluino ($\tilde g$), the chargino ($\tilde \chi_1^{\pm}$), the neutralino ($\chi_1^0$), the stau ($\tilde \tau_1$),
  and the stop ($\tilde t_1$) are exhibited in the inset where the y axis is plotted on a logarithmic scale.
\label{reachplot} }
\end{center}
\end{figure}

 A  number of works already  exist which analyze the signatures for supersymmetry at $\sqrt s=14$ TeV
 and $\sqrt s=10$ TeV (for a small sample see~\cite{Hubisz:2008gg,Baer:2008ey,Dietrich:2010tf,14tev}). {In our analysis we computed the sparticle branching ratios with SUSY-HIT~\cite{Djouadi:2006bz} and they were subsequently read into Pythia.  PGS~4 was used for the detector simulation{ with no trigger imposed (L0)}. }
 Here we give the analysis for $\sqrt s=7$ TeV  {for} a large number of signatures for each candidate  model
 considered.
 These are  listed in Table~(\ref{SUSYcuts}) and consist of a combination of multijets,  b-tagged jets,
 multileptons,   jets and leptons,
   and photons  with a variety of cuts  geared to reduce the standard model background and enhance
 the SUSY signal with and without missing energy.
We discuss first the model with universal soft
breaking, i.e., the mSUGRA model.
 In Fig.~(\ref{reachplot})  we give the reach of the early runs at the
LHC for mSUGRA in the { $m_{1/2}-m_{0}$} plane.  One finds that the reach   can extend
 up to about 400 GeV for  $m_{1/2}$ at low values of $m_0$ and up to about  2 TeV for $m_0$
 for low values of $m_{1/2}$ with 1~fb$^{-1}$ of integrated
luminosity, and up to about  450 GeV for 2~fb$^{-1}$ of data, and for $m_0$ the reach can extend up to
1.9 (2) TeV for 1(2)fb$^{-1}$ of integrated luminosity.
For the nonuniversal case a plot in $m_{1/2}-m_{0}$ is not very illuminating because of the
presence of nonuniversalities. Here  we exhibit
in Table~(\ref{bench2})  a set of benchmarks
 which have a chance of early discovery. A criterion used in the selection of these benchmarks is the size
 of the cross section for the production of SUSY events $\sigma_{SUSY}$ which is shown in the last column of
 Table~(\ref{bench2}).
Here one finds that $\sigma_{SUSY}$ for some of the benchmarks is as large  as  10-20 pb or more implying that as
many as $(1-2)\times 10^{4}$ SUSY events will be produced at the LHC with  1~fb$^{-1}$ of integrated
luminosity. Thus with efficient cuts to reduce the SM background there appears a good chance for the
discovery of such models. A detailed analysis of the signatures
implementing the cuts of   Table~(\ref{SUSYcuts})
bears this out.
Thus as exhibited in Fig.~(\ref{grid}) one finds that all of the
benchmarks  of Table~(\ref{bench2})  do indeed produce visible  signals not just in one but in several channels.
In fact  for { most} the benchmark models of  Table~(\ref{bench2})
 one has as many as five channels  and often more
   where the SUSY signal will become visible, thus providing  important cross-checks
for the discovery of supersymmetry.
In Fig.~(\ref{grid}) we also exhibit discovery channels for   0.1~fb$^{-1}$, 1~fb$^{-1}$, and 2~fb$^{-1}$ of data
which shows the {identities of} the new signature channels that open  as one increases the {integrated} luminosity.
 It is interesting to ask how the number of visible signatures depends
on the integrated luminosity. An analysis of this issue is given in Fig.~(\ref{lum}) where a plot is given
of the number of signature channels where the SUSY signal becomes visible as a function of  the  integrated luminosity.
The figure shows that the number of discovery channels increases rather sharply with luminosity
and can become as large as 10 or more
 at 1~fb$^{-1}$ of integrated luminosity. Quite interestingly  the
analysis of Fig.~(\ref{lum})  also exhibits that a SUSY discovery can occur with an integrated luminosity as
low as 0.1~fb$^{-1}$ still with several available discovery channels.

\vspace{.1cm}
 {\tiny{
  \begin{figure}[!t]
{\bf A Display of Visible Discovery Channels for 0.1 fb$^{-1}$, 1 fb$^{-1}$ and 2 fb$^{-1}$ at $\sqrt{s}=7\text{ TeV}$.}
\begin{center}
\includegraphics[scale=0.55]{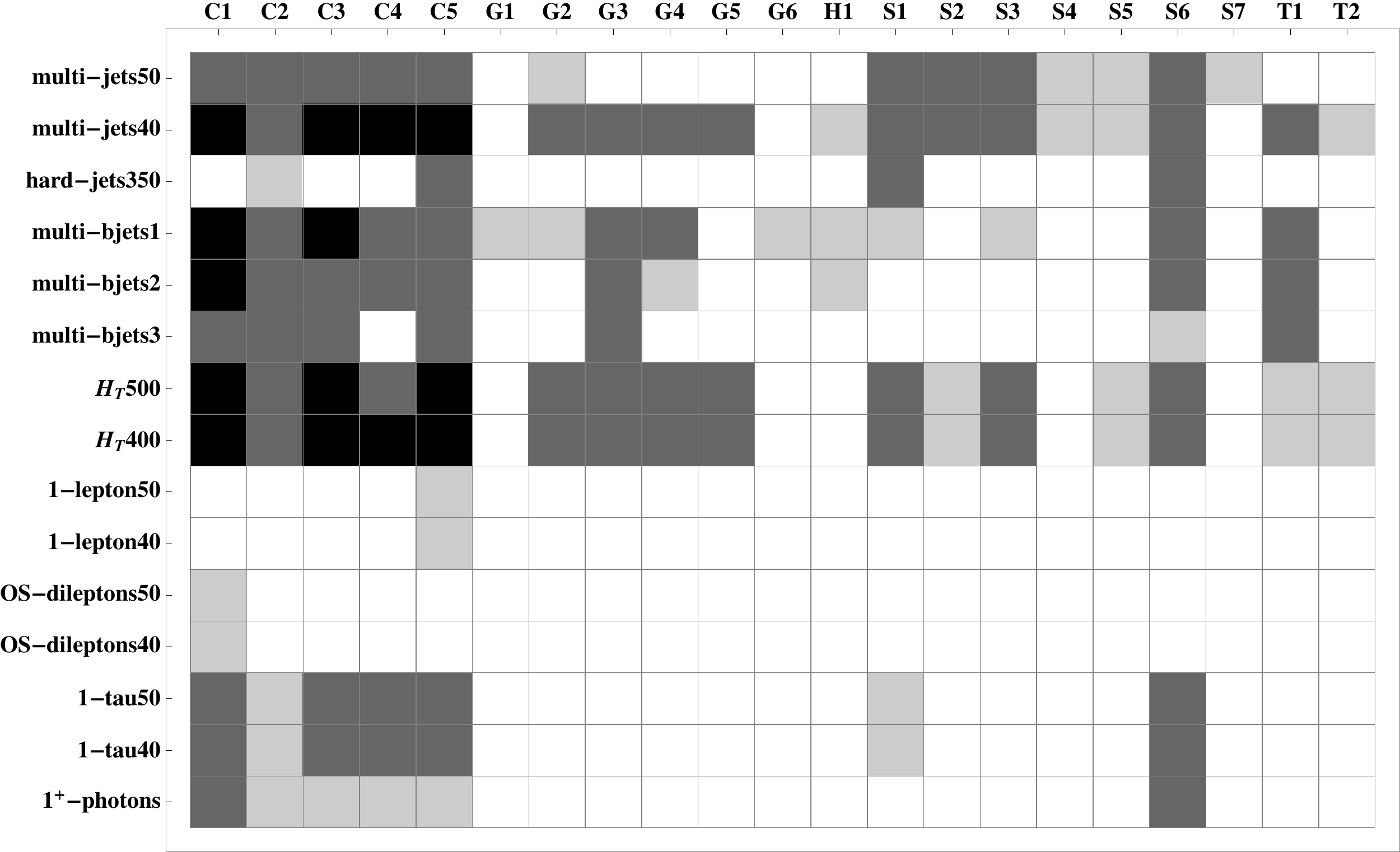}
\caption{{An exhibition of the visible discovery channels for 0.1 fb$^{-1}$ (black squares), 1 fb$^{-1}$ ({dark gray squares}) and 2 fb$^{-1}$ ({light gray squares}) at $\sqrt{s}=7\text{ TeV}$.
The discovery channels are listed in Table~\ref{SUSYcuts}. {  Here we are using the convention that C denotes a chargino model, G denotes a gluino model, H denotes a Higgs model, S denotes a stau model, and T denotes a stop model.}}
\label{grid}}
\end{center}
\end{figure}
  }}
 \begin{figure}[!h]
{\bf A Display of the Number of Discovery Channels as a Function of Integrated Luminosity.}
\begin{center}
\includegraphics[scale=0.8]{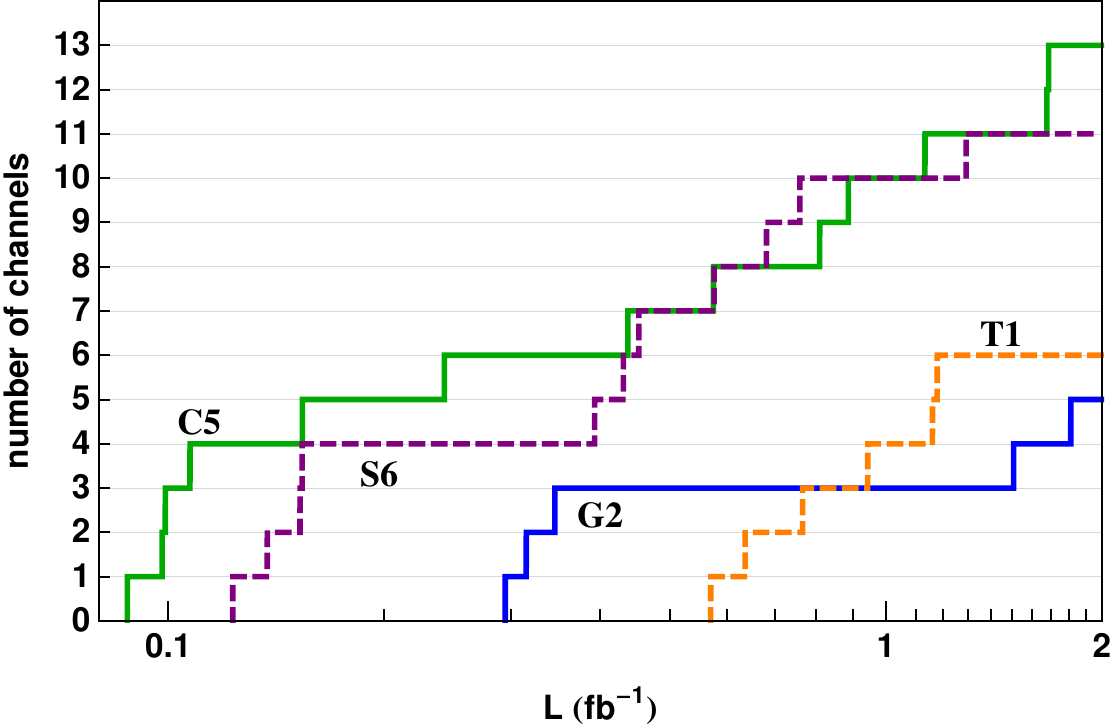}
\caption{{An exhibition of the rapid rise in the number of discovery channels  vs integrated luminosity
for four early discovery benchmarks  given in Table~(\ref{bench2}).
The number of discovery channels  for supersymmetry in each case is  in excess of five and in some cases as large
as 10 or above  at  $1\text{ fb}^{-1}$ of data at $\sqrt s=7$ TeV.}
\label{lum}}
\end{center}
\end{figure}
\begin{figure}[!h]
{\bf A Display of Benchmarks as Dark Matter Candidates.}
\begin{center}
\includegraphics[scale=0.8]{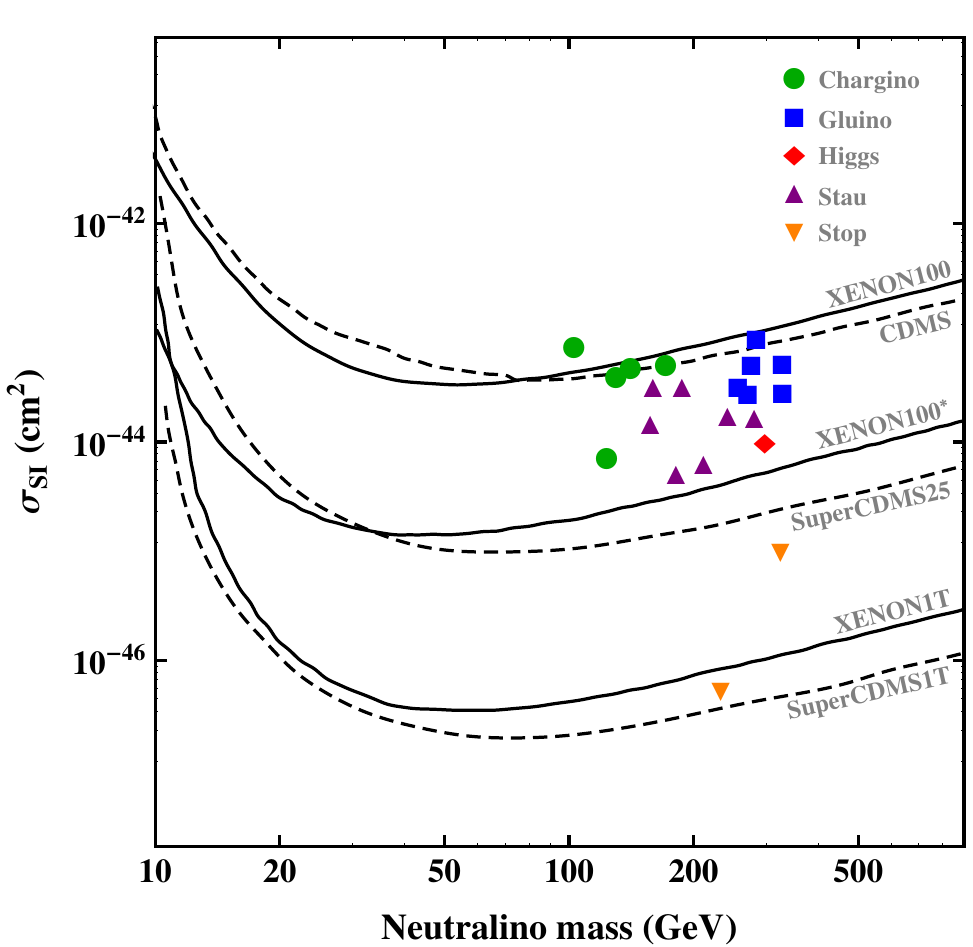}
\caption{\label{SIxsect}An exhibition of the spin independent {neutralino-proton} cross section{, $\sigma_{\text{SI}}$,} for the {benchmark} models.  These  are labeled by the NLSP, which under the applied constraints allow chargino (green circles), stau (purple triangles), gluino (blue squares), CP odd Higgs (red diamond), and stop (orange inverted triangles) NLSPs.   In the plot the curve labeled XENON100* is the expected sensitivity
of XENON100 with $6000kg\times days$ of data,  the curve labeled XENON1T is the expected sensitivity  for $1 ton \times year$ of data and the curves labeled  SuperCDMS25 and SuperCDMS1T
are  the expected sensitivities for  the  two SuperCDMS experiments.}

\end{center}
\end{figure}

As noted earlier all of the benchmarks listed in Table~(\ref{bench2}) are discoverable and further, as shown in
Fig.~(\ref{SIxsect}), all of  them are also consistent with the current limits on the spin independent {neutralino-proton}
cross sections from CDMS-II and XENON-100 {and can be seen in the next generation of xenon and germanium experiments}.
Finally in Table~(\ref{susy})
we exhibit  the light sparticle spectrum for a subset of the  benchmarks given in Table~(\ref{bench2}).
Here  we note that
some of the models in Table~(\ref{bench2}),
as, for example, the model {C1},
  have typically a small value of $\mu$ indicating that they reside on or near the so-called hyperbolic branch/focus point region~\cite{Chan:1997bi}  of radiative breaking of the electroweak symmetry.
  Such models would have  scalar  masses which are  typically  heavier and often much heavier than the gaugino masses as can be seen
  in Table~(\ref{susy}).

\section{Model simulations\label{modelSim}}
We give now a further discussion of the model simulations. As
Fig.~(\ref{grid}) indicates, the primary discovery channels for
supersymmetry at $\sqrt{s}=7\TeV$ will be jet-based signatures which
are designed to be as inclusive as possible to increase the number
of signal events. This will need to take precedence over signal
purity (i.e. efficiency to reject background) at values of
integrated luminosity at or below 1~fb$^{-1}$.
Four of the five chargino NLSP benchmarks can be discovered via
jet-based signatures within the first 100~pb$^{-1}$ of data, with
the remainder (Chargino 2) reaching a five-sigma significance in the
{multijets100} channel within 200~pb$^{-1}$. The stau NLSP models
favor traditional multijet signatures such as {multijets200} and
$H_T\,500$ which involve much harder jet-$p_T$ requirements. Among
our benchmarks the stau NLSP cases tend to have the heaviest gluinos
with $709\GeV \leq m_{\tilde{g}} \leq 912\GeV$ and lightest squarks
in the range of 330-600 GeV. These examples will therefore exhibit
very similar characteristics in the early data taking, with the
heavier gluino producing long cascades of moderately energetic jets
which satisfy the high-$p_T$ jet requirements common in analyses at
$\sqrt{s}=14 \TeV$. By contrast, our Higgs, stop, and gluino NLSP
models favor discovery channels with much looser jet requirements to
increase the signal size.  These include the $H_T\,400$,
multi-bjets1 and multi-bjets2 channels. We can understand the
prevalence of b-jet signatures because of the rather small mass gaps
between the lightest $SU(3)$-charged state (i.e. the gluino or
squark) and the LSP which eventually appears at the end of the
cascade~\cite{Feldman:2009zc}. For example, the mass gap between
the gluino and the LSP for the gluino NLSP models ranges from 60 to
160~GeV, with an average gluino mass for these models of 390~GeV. We
note that these b-jet channels are defined with a requirement that
$p_T^{\rm jet}\geq 40\GeV$ for the b-tagged jet and we impose a veto
on isolated leptons ($e$ and $\mu$). The latter is not strictly
necessary to achieve a high signal significance, but imposing the
leptonic veto sufficiently reduced the SM background arising from
the $t$,$\bar{t}$ background samples to increase signal significance
by a factor of about 65\% on average.

The discussion above indicates an important concern for
experimentalists facing the challenge of extracting the most signal
possible from the first year of LHC data. Traditional signature
definitions are often designed to enhance the signal-to-background
purity so as to be able to make meaningful exclusive measurements of
key quantities -- often the edges or end points of various kinematic
distributions. As we will see below, there will likely be few
opportunities to employ these analysis techniques with only
1~fb$^{-1}$ of data at $\sqrt{s}=7\TeV$, even if we are fortunate
enough to discover supersymmetry in multiple channels. It thus may
make sense to seek, {at the expense of purity,} a more inclusive set
of signature definitions. {Consider, for example,
the object defined by
\begin{equation}
\displaystyle\sum_{i=1}^4 p_{Ti}+\slashed{E} \, .\label{HTeqn}
\end{equation}
When the four objects entering~(\ref{HTeqn}) are restricted to the
four hardest jets, the above is typically referred to as effective
mass ($M_{\rm eff}$), and has often been studied in the context of
multijet channels for discovery of
supersymmetry~\cite{Hinchliffe:1996iu,Baer:1995nq}. When we wish to restrict our
attention to jets only, we will refer to~(\ref{HTeqn}) as the ``jet
effective mass'' of the event.
A more inclusive definition is to allow the four hardest visible
objects to enter~(\ref{HTeqn}) -- in particular, hard leptons. For
the sake of clarity, when we wish to consider this more expansive
variable we will refer to it as ``generalized effective mass'' or
$H_T$, as we did in Table~\ref{SUSYcuts}. In the case of this
generalized effective mass we will also impose an} overall
collective cut on the scalar sum of all $p_T$'s for these objects,
as opposed to cuts on individual object $p_T$'s. In an environment
where jet-$p_T$ measurements are likely to be less accurate than we
might like, this is a reasonable variable to employ and we find it
to be particularly effective across many of our benchmark
categories.
Similarly, we have given in Table~(\ref{SUSYcuts}) a number of
variants to the jet-$p_T$ requirements to give the reader the sense
of how relaxing these requirements may accelerate discovery. We will
investigate this further with a few of our benchmarks below. In
general, the lighter jet requirements tend to result in a signal
which becomes significant sooner, but when both channels are
significant the stronger jet case may often have the better
signal-to-background figure.

We find very few discovery channels in the leptonic sector for our
benchmark models. In the one-lepton plus jets channels at
1~fb$^{-1}$ we find between 10 and 50 events for our chargino NLSP
and stau NLSP models and less than 10 events for the Higgs, gluino
and stop NLSP models. This is to be contrasted with 367 background
events with the 40~GeV jet requirement (1-lepton40) and 350 with the
stricter jet requirement (1-lepton100). These channels achieve five-sigma significance only for model Chargino~5, and only after
2~fb$^{-1}$ of data. The opposite-sign (OS) dilepton channels give
$\mathcal{O}(10)$ signal events after 1~fb$^{-1}$ of data for our
chargino NLSP models only. These cases just fail to reach five-sigma
significance until 2~fb$^{-1}$ of data. Event rates for same-sign
dileptons and trileptons are consistent with zero for all benchmarks
at 1~fb$^{-1}$, with a handful of events expected at 2~fb$^{-1}$.
These ``gold-plated'' supersymmetry channels will need to wait for
higher center of mass energies or larger data sets to make their
presence manifest.

Finally, we make a few remarks about discovery channels with an
imperfect detector. Early results from both detectors at the LHC
seem to indicate that measurement of missing transverse energy
appears to fit Monte Carlo predictions within reason and the overall
error in determining missing $E_T$ is less severe than many
pessimists had feared~\cite{met}. This is welcome news, as our
benchmarks will all require missing $E_T$ as part of any search
strategy. None of the benchmarks we studied are visible above
backgrounds without this important ingredient, with the number of
signal events becoming comparable to the SM background only when
$\slashed{E}_T \geq 200\GeV$ or higher is employed.
We also point out that many of our best signals involve tagged
b jets. As mentioned previously, we updated the PGS4 b tagging to
reflect the more optimistic expectations of the ATLAS and CMS
collaborations. Achieving these efficiency targets early in LHC
data taking will be important for many supersymmetric models
relevant to the LHC at $\sqrt{s}=7\TeV$.

To understand these features we will look a bit more closely at
eight of our benchmarks from Table~(\ref{bench2}) chosen to
represent an array of phenomenologies at the LHC. These eight
benchmarks are collected in Table~(\ref{susy}) where we provide
physical masses of some of the key superpartners relevant for LHC
signatures. For each benchmark we give in Table~(\ref{discovery})
the number of signal events in 1~fb$^{-1}$ for a selection of
signatures from Table~(\ref{SUSYcuts}), as well as the standard
model background count and the signal significance. In this table we
have included, for illustrative purposes, a modification of the
multi-jet signature in which we merely require at least four jets,
each with $p_T^{\rm jet} \geq 40 \GeV$ in addition to the missing
transverse energy and transverse sphericity cut. The table
illustrates the relative paucity of leptonic signatures across all
of our benchmark models and the importance of employing relative
loose jet-$p_T$ requirements in multijet signatures.

\begin{figure}[t]
  \begin{center}
{\bf Jet Effective Mass and Missing Transverse Energy
Distributions}\\ \centering
 \includegraphics[width=8cm,height=6cm]{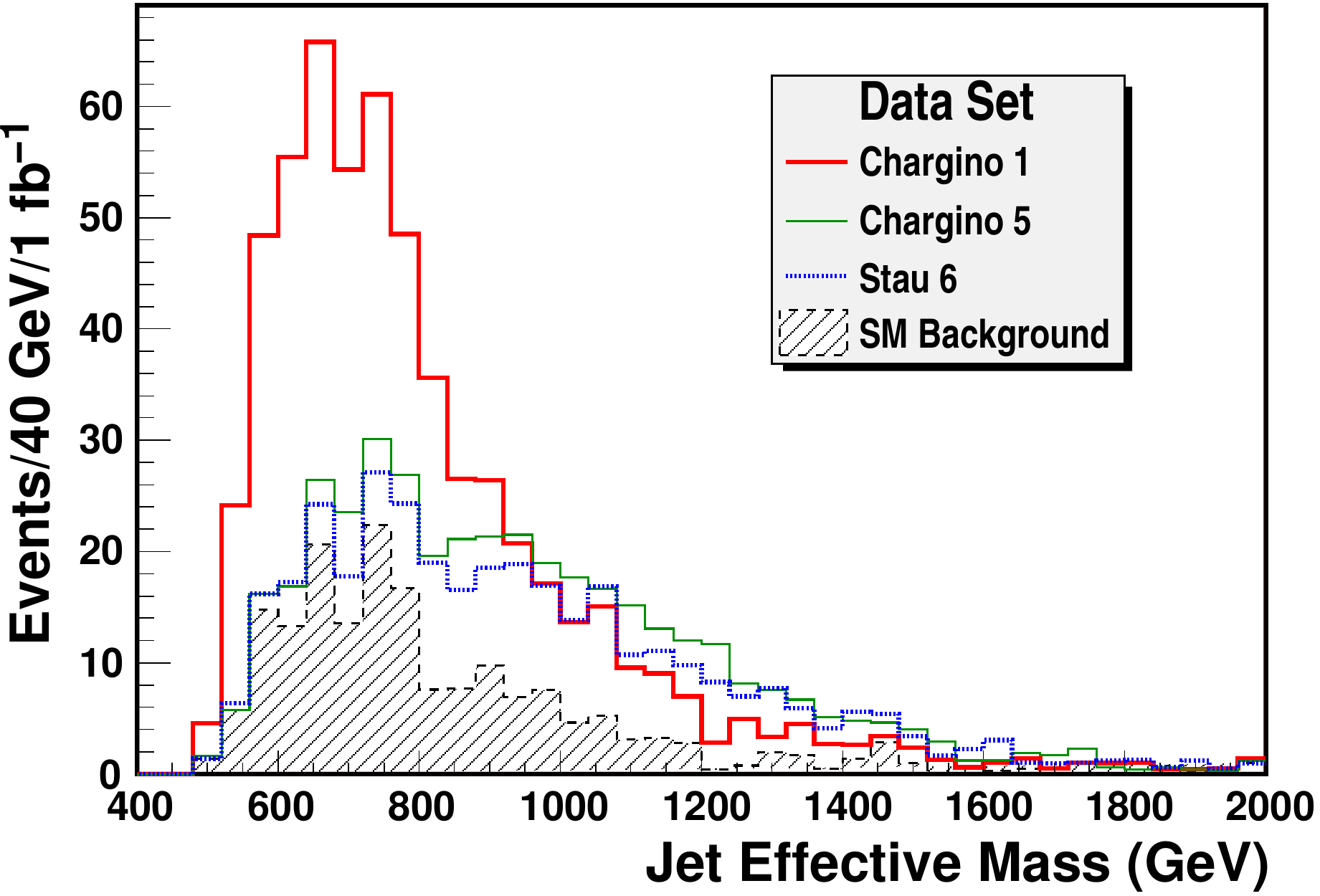}
 \includegraphics[width=8cm,height=6cm]{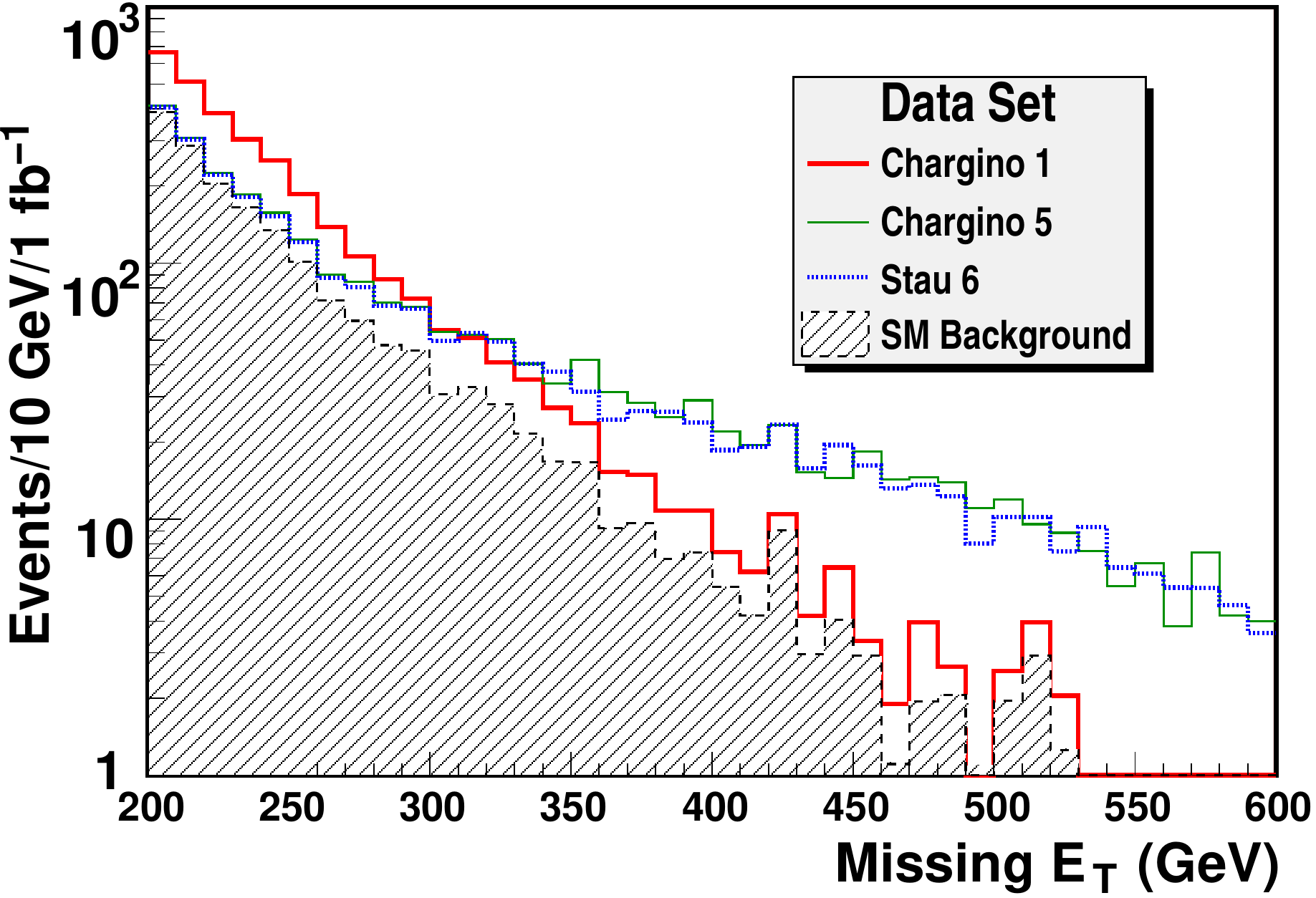}
 \caption{Left: Effective mass distribution (multi-jets100) of the
 four hardest jets for three benchmark models with strong signals in
 this channel: Chargino~1 red (heavy) line, Chargino~5 green (thin) line and
 Stau~6 blue (dash-dotted) line. This signature is defined in Eq. {(\ref{HTeqn})} in the
 text. Signal curves are signal + SM background, with the background
 component indicated by the hatched region. Right: Missing
 transverse energy distribution for the same three models on a
 logarithmic scale. Again the SM component is indicated by the
 hatched region.}
    \label{graphmeff}
  \end{center}
\end{figure}

In the left panel of Fig.~(\ref{graphmeff}) we exhibit the
multijets100 signature, defined by { Eq. (\ref{HTeqn})}, for the three
benchmarks of Table~(\ref{discovery}) with the strongest signal. The
distribution of $m_{\rm eff}$ for the three models is given in
terms of the signal plus the SM background, normalized to
1~fb$^{-1}$, with the SM background also shown separately in the
hatched region. The strongest signal comes from Chargino~1 whose
overall distribution is very similar in size and shape to that of
the SM background, while Chargino~5 and Stau~6 peak at slightly
larger values. These distributions are heavily influenced by the
distributions in $\slashed{E}_T$, which are plotted on a logarithmic
scale in the right panel of Fig.~(\ref{graphmeff}). We can
understand these shapes from the values in Table~(\ref{susy}): the
relatively light gluino of model Chargino~1 is the lightest
$SU(3)$-charged state which will decay to low-$p_T$ jets, while the
lightest such state in the other two cases is the stop which has a
slightly higher mass ($m_{\tilde{t}} = 402\GeV$ for Chargino~5 and
497~GeV for Stau~6). For these two cases the dominant supersymmetric
production processes involve the associated production of a gluino
and a squark, with the decay of the gluino to on-shell squarks
producing slightly harder jets. An examination of the multijet
$m_{\rm eff}$ variable with varying jet requirements should reveal
the general mass scale for the low-lying strongly coupled
superpartners, and this information should { assist in optimizing} jet-$p_T$
requirements on signals with lower statistics.

\begin{figure}[t]
  \begin{center}
{\bf Changing Jet-$p_T$ Requirements on Multi-Jet Signatures}\\
\centering
 \includegraphics[width=8cm,height=6cm]{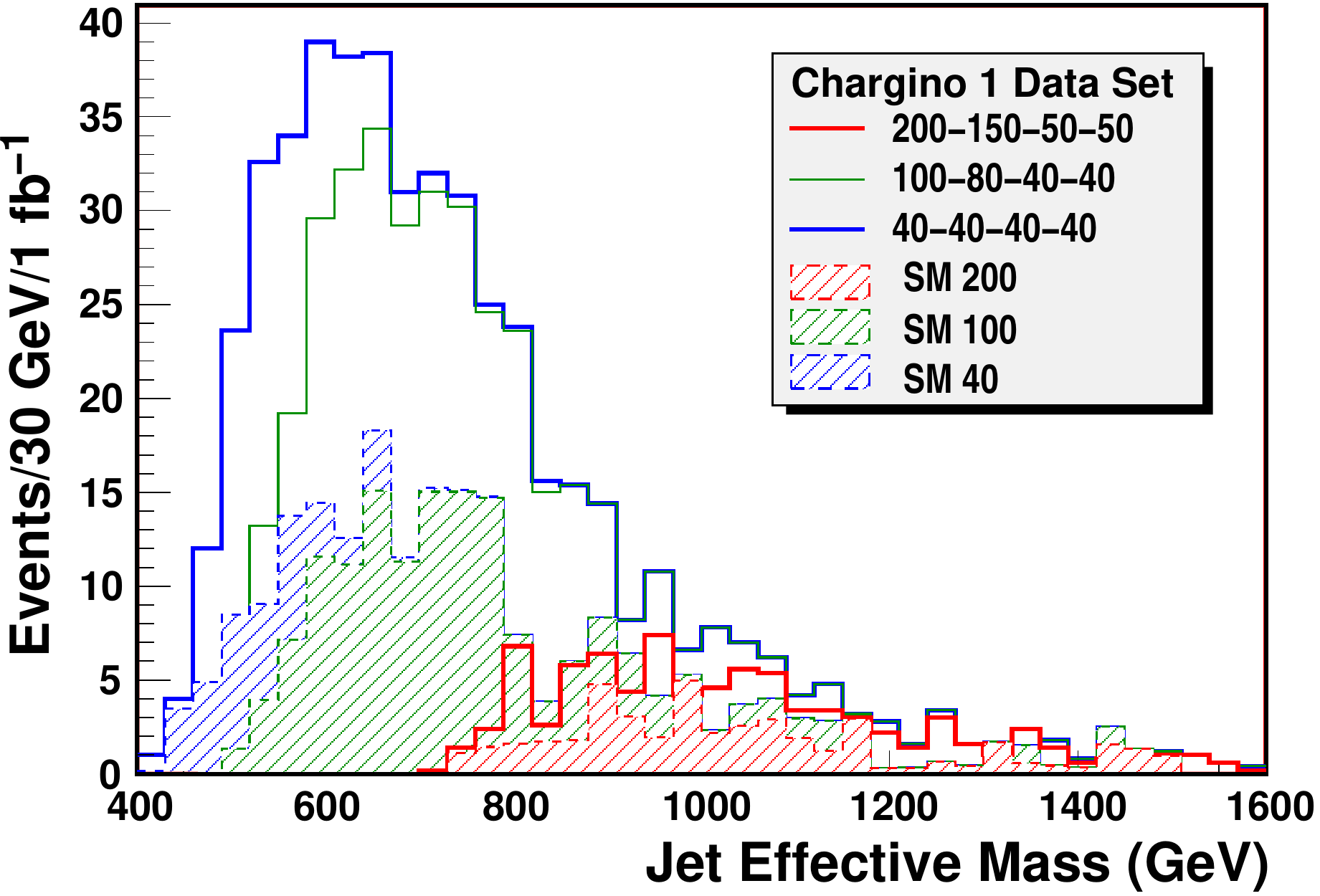}
 \includegraphics[width=8cm,height=6cm]{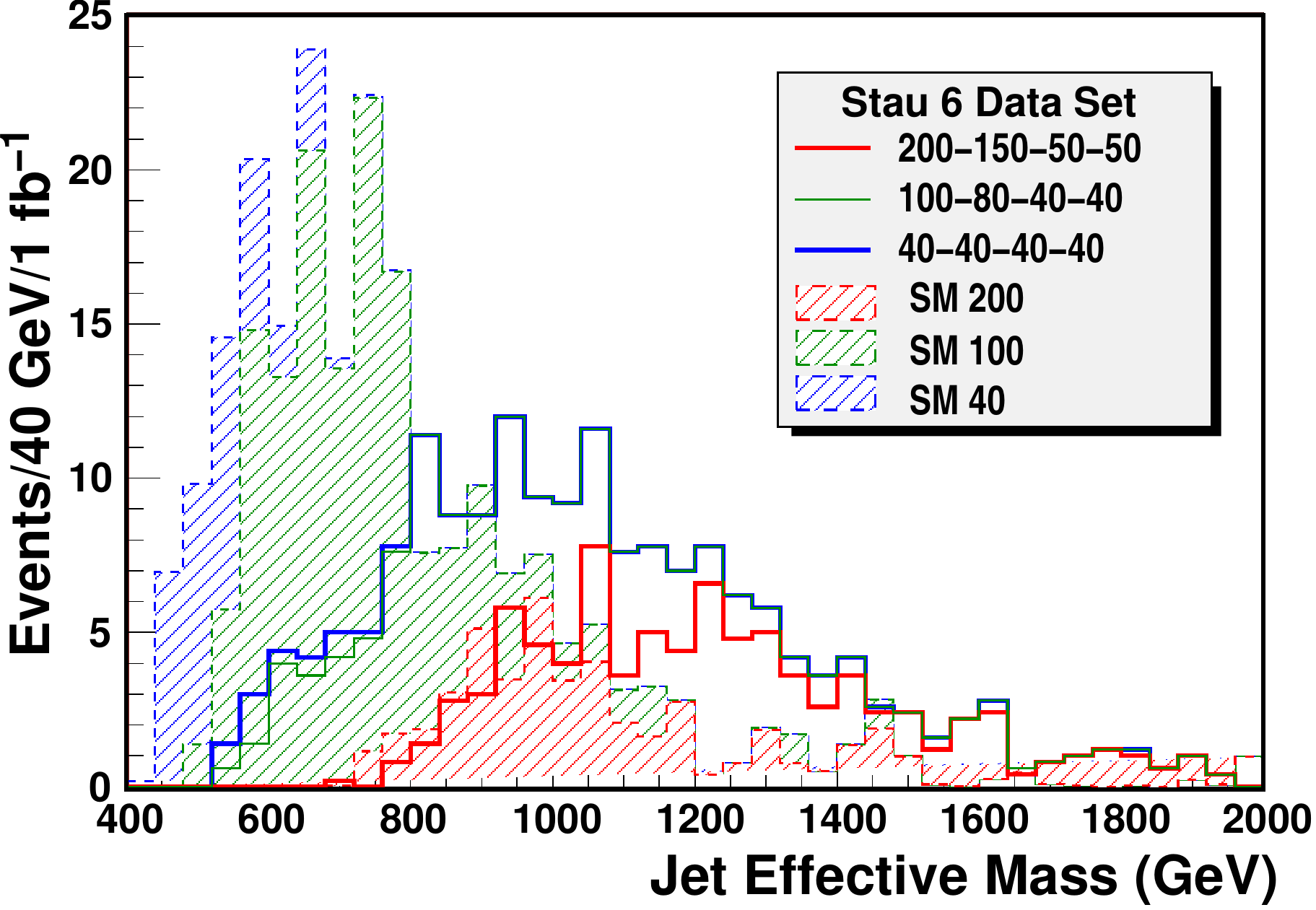}
 \caption{The effect of changing the minimum transverse momentum of
 leading jets for multijet signatures is exhibited for two models: Chargino~1
 (left panel) and Stau~6 (right panel). Heavy (red line) distributions
 represent the strict multijets200 signature, thin (green line)
 distributions are the softer multijets100 signature, and medium
 (blue line) distributions are for the multijets40 signature. The legend
 gives the minimum $p_T^{\rm jet}$ requirement for the four hardest
 jets in the event. In this figure we
show the signal superimposed on the SM background separately to
distinguish between the three cases.}
    \label{graphcuts}
  \end{center}
\end{figure}

In Fig.~(\ref{graphcuts}) we investigate the effect of changing
these jet-$p_T$ requirements for Chargino~1 (left panel) and Stau~6
(right panel). Here we include the signatures denoted multi-jets200
and multi-jets100 in Table~(\ref{SUSYcuts}) as well as our softer
signature multi-jets40 from Table~(\ref{discovery}). In this case we
show the signal superimposed on the SM background separately to
distinguish among the three cases. Loosening the jet requirements
populates the lower energy bins in $m_{\rm eff}$ for both the signal
and the background, leaving the higher energy bins mostly
unaffected. Therefore these softer cuts tend to boost the signal
significance of models such as Chargino~1 with softer jets, while
actually reducing the significance of models like Stau~6 where jets
carry much more $p_T$ on average. This is also reflected in the
signal significances at 1~fb$^{-1}$ in Table~(\ref{discovery}). For
these models all three signature definitions yield a discovery, but
for other models such as Gluino~2 the softer jet requirement is
crucial to making an early discovery. In Fig.~(\ref{graphht}) we
plot the generalized $H_T\,400$ variable for the three models for
which this is the leading discovery channel: Chargino~5, Stau~6, and
Gluino~2. For the last case, this channel produces a discovery in
approximately 300~pb$^{-1}$ while other channels must wait for more
data for a five-sigma excess. Despite the very light gluino in model
Gluino~2, the compressed spectrum of this model produces extremely
soft jets and a relatively low jet multiplicity for gluino pair
production events. Expanding the definition to include all visible
decay products produces a stronger signal.

\begin{figure}[t]
  \begin{center}
{\bf Generalized $H_T$ Signature}\\
\centering
 \includegraphics[width=8cm,height=6cm]{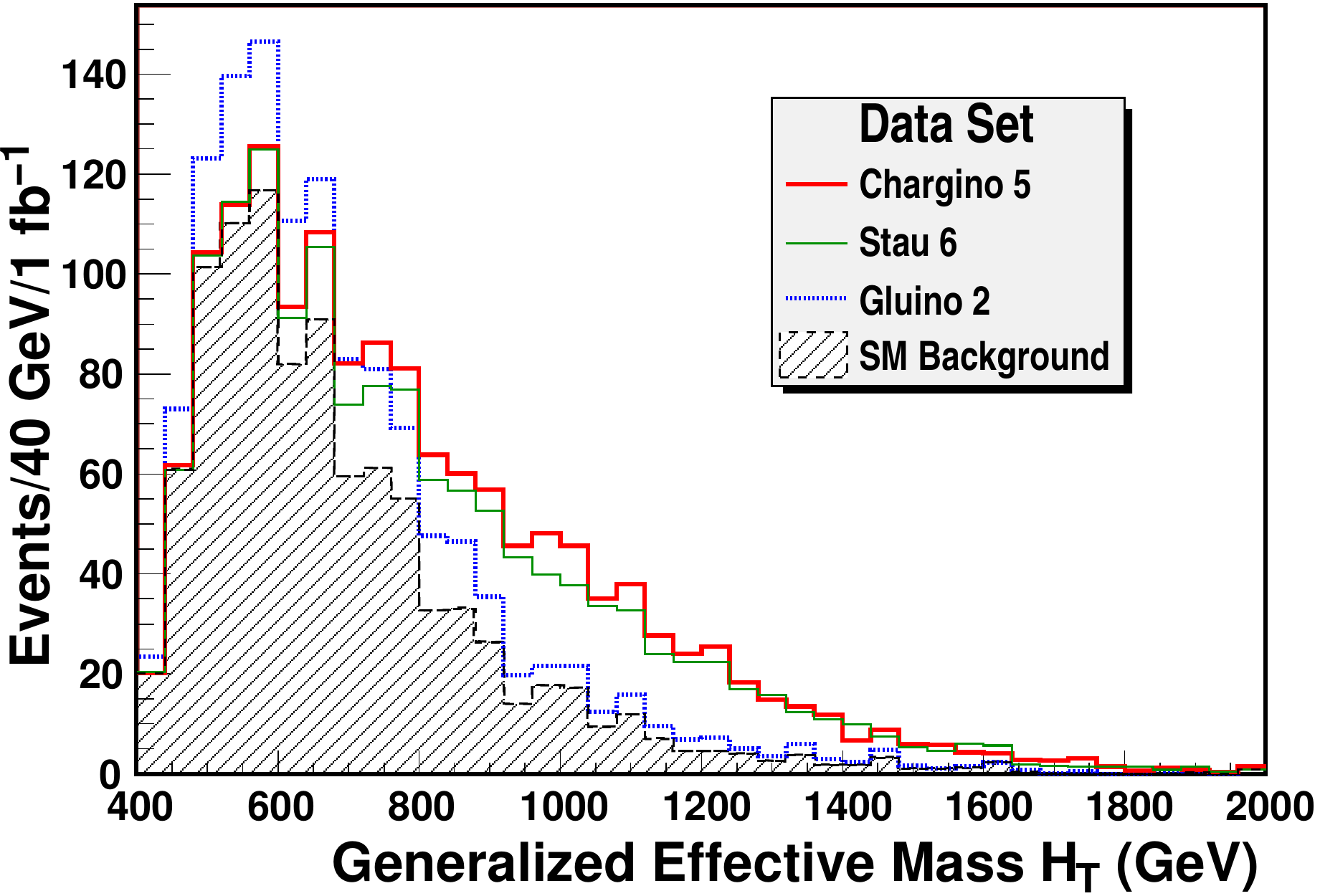}
 \caption{Distribution of $H_T$, defined by { Eq. (\ref{HTeqn})} but applied to the
 hardest four visible objects in the event as opposed to the four hardest jets with
 an isolated lepton veto. The signal  plus background distribution is given for the
 three benchmarks for which this was the best discovery channel:
 Chargino~5 heavy (red) line, Stau~6 thin (green) line and Gluino~2 medium (blue)
 line. The component of the distribution made up of SM background events is indicated
 by the hatched region.}
    \label{graphht}
  \end{center}
  \vspace{-0.5cm}
\end{figure}
\begin{figure}[h!]
  \begin{center}
{\bf Opposite-Sign Dilepton Invariant Mass Distribution}\\
\centering
 \includegraphics[width=8cm,height=6cm]{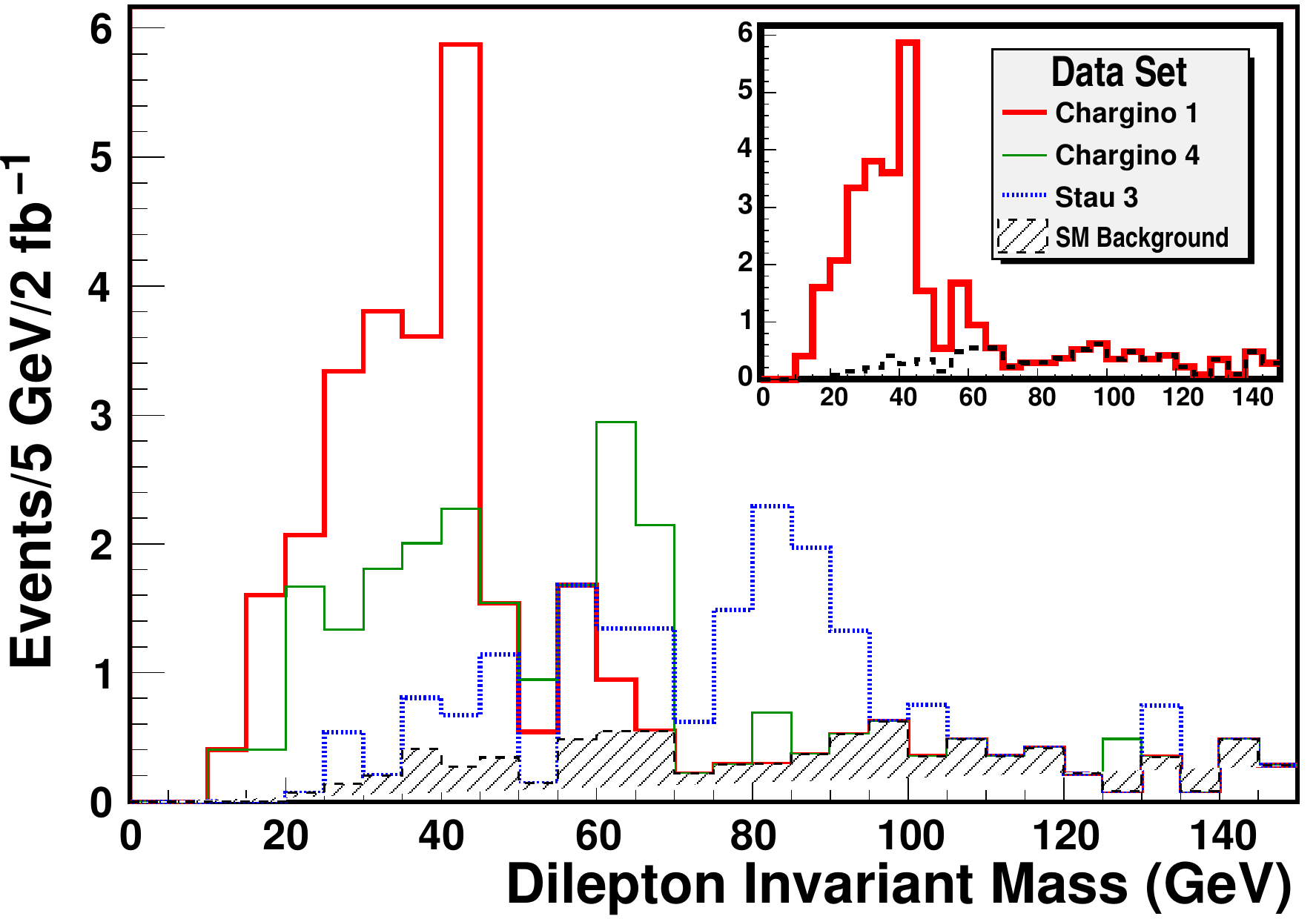}
 \caption{The opposite-sign dilepton invariant mass is displayed for events
 satisfying the signature definitions denoted OS-dileptons40 in Table~(\ref{SUSYcuts}).
 The models shown are Chargino~1 heavy (red) line, Chargino~4 thin (green) line
 and Stau~3 medium (blue) line. As with previous distributions the curves are for
 the signal + SM background, with the background component indicated by the hatched region.
 The inset shows the distribution for Chargino~1 alone, to exhibit the edge at
 approximately 50~GeV.}
    \label{graphdilep}
  \end{center}

\end{figure}

As mentioned above, the peak in the jet effective mass distribution
-- or the energy at which the signal begins to exceed the SM
background -- is a relatively good indicator of the rough mass scale
of the lightest $SU(3)$-charged
superpartner~\cite{Hinchliffe:1996iu}. Apart from this rough estimate, {one might ask if there are other properties of the superpartner spectrum that can be ``measured," even at such low statistics.} We believe that the
answer is in principle yes, depending on the model. The most
well-known measurement technique is to form the invariant mass of
opposite-sign dilepton pairs and look for an edge, or end point, in
the distribution. This procedure is often performed after a
flavor subtraction is done (i.e., the combination
$e^+e^-+\mu^+\mu^--e^+\mu^--e^-\mu^+$ is formed) to reduce SM and
SUSY combinatorial backgrounds. After 1~fb$^{-1}$ there is unlikely
to be sufficient numbers of OS dilepton events to perform this
subtraction, or even to identify a true edge. Even after 2~fb$^{-1}$
the statistics are sufficiently low to make this measurement
difficult in all but the most favorable model points. We illustrate
this in Fig.~(\ref{graphdilep}), normalized to 2~fb$^{-1}$ of
data. It is tempting to see an edge in the invariant mass
distribution for Chargino~1 (see inset for a clear view). But the
low statistics are evidenced by the scale on the vertical axis. The
sharpness of the edge is largely an artifact of simulating our
signal sets at the 5~fb$^{-1}$ {level} and then rescaling to a smaller
integrated luminosity. Nevertheless, this edge is indeed real and
accurately describes the mass distribution between the lightest and
second-lightest neutralino. The first hint of the ``spoiler mode''
$\chi_2^0 \to Z\,\chi_1^0$ present in the Stau~3 benchmark is also
present in Fig.~(\ref{graphdilep}).

Potentially more promising are measurements based on the invariant mass distribution of events with precisely two b-tagged jets.  In  Fig.~(\ref{graphbjet}) we give di-bjet invariant mass distributions for four models with particularly strong signals in this channel.  In the right panel we give two models for which the b-jet channels are the strongest discovery modes, Gluino 3 and Stop1, and with a sufficient amount of data it should be able to measure the mass difference between the gluino and the lightest neutralino.

\begin{figure}[t]
  \begin{center}
{\bf Invariant Mass Distribution of B-Jets}\\
\centering
 \includegraphics[width=8cm,height=6cm]{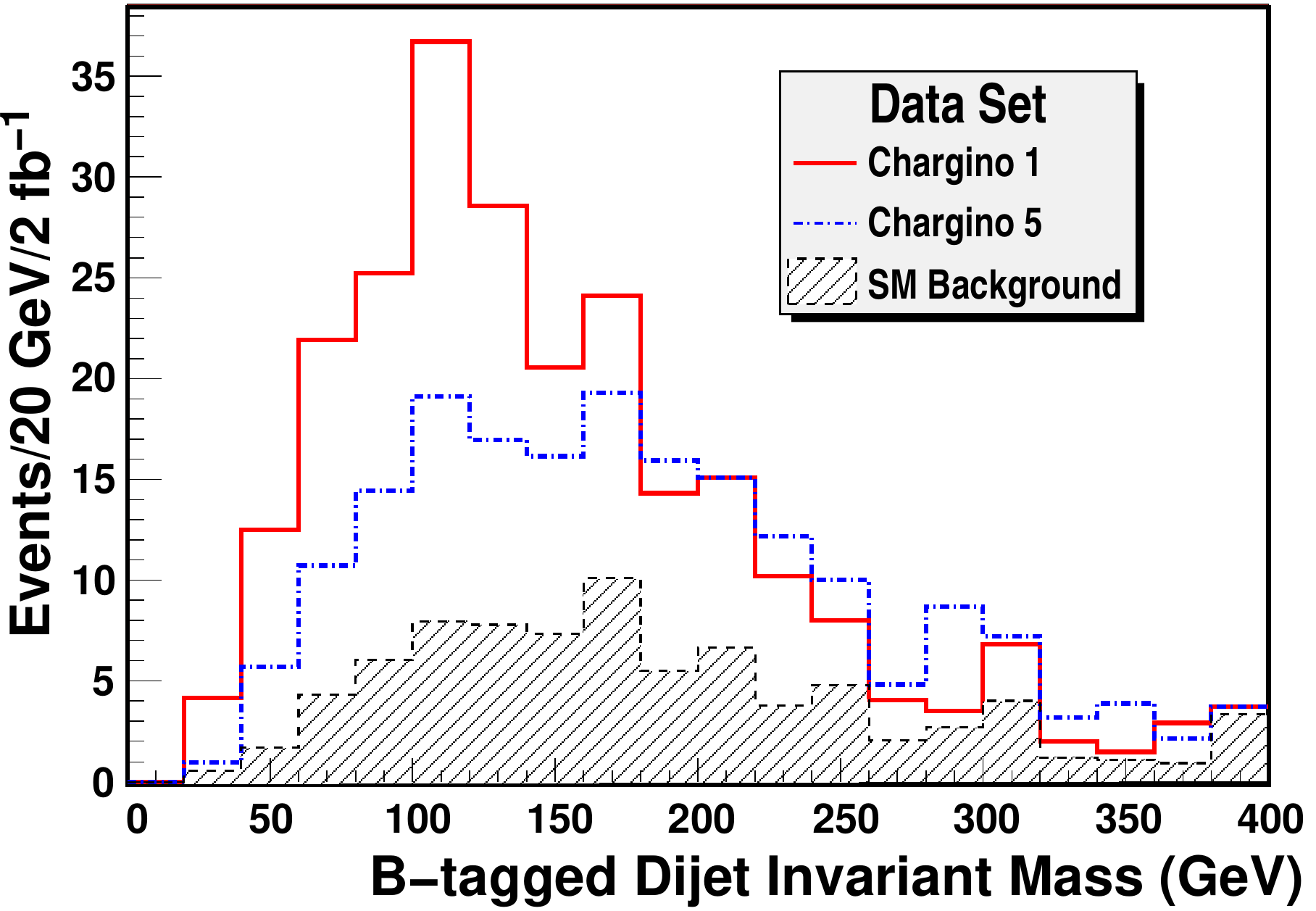}
 \includegraphics[width=8cm,height=6cm]{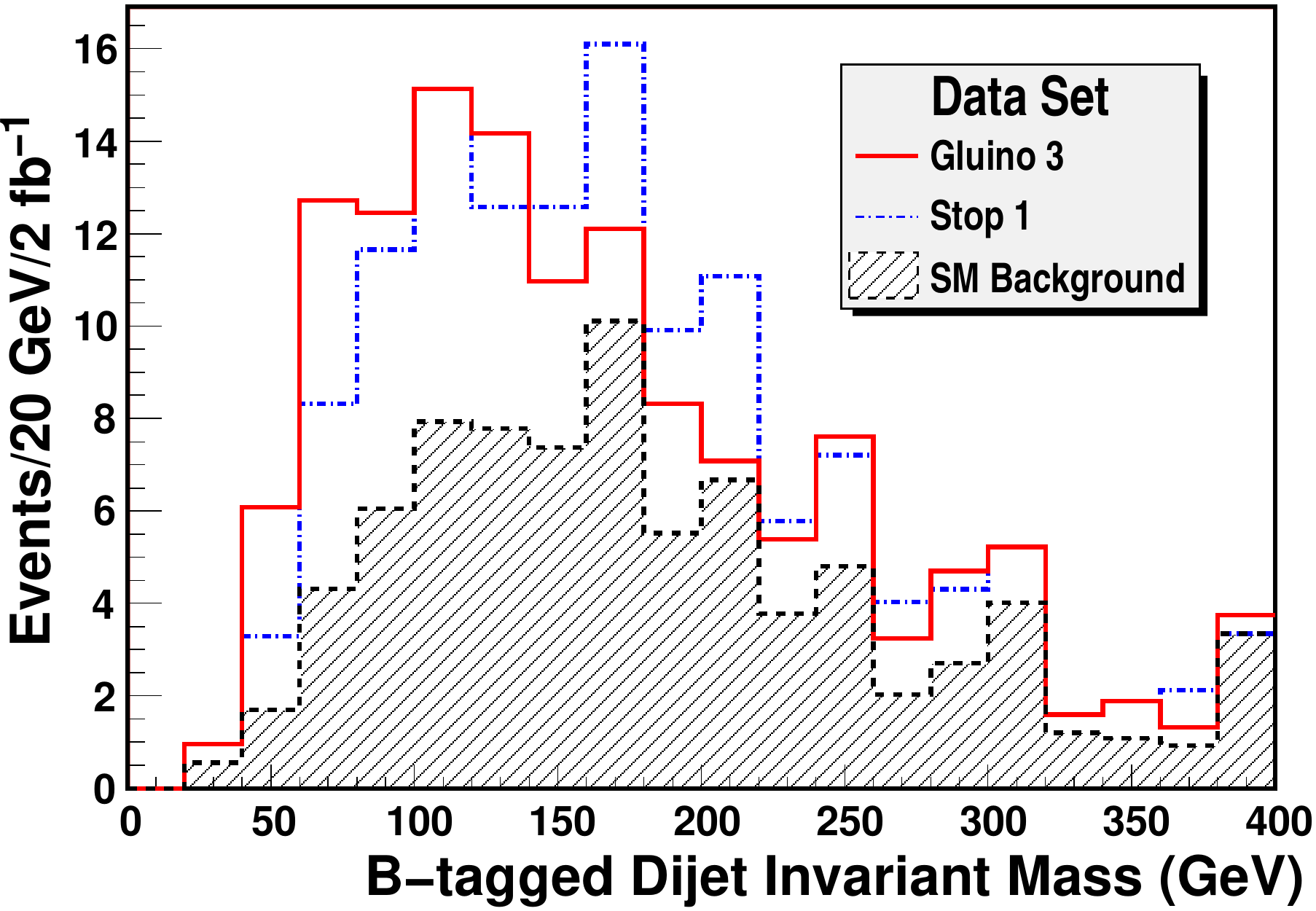}
 \caption{The invariant mass distribution of b-jet pairs for events with
 precisely two b-tagged jets, each with $p_T^{\rm jet} \geq 40 \GeV$,
 is given for events which satisfy our initial cuts of transverse sphericity
 $S_T \geq 0.2$ and at least 200~GeV of $\slashed{E}_T$. No lepton
 veto is imposed on these distributions. As with previous distributions the curves are for
 signal + SM background, with the background component indicated by the hatched region. Left panel: Chargino~1
 heavy (red) line and Chargino~5 dash-dotted (blue) line. Right
 panel: Gluino~3 heavy (red) line and Stop~1 dash-dotted (blue)
 line.  }
    \label{graphbjet}
  \end{center}
  \vspace{-.5cm}
\end{figure}

\section{Conclusions\label{conclusion}}
We have given here an analysis of the potential of the LHC in early runs at $\sqrt s=7$ TeV to discover
 supersymmetry with 1~fb$^{-1}$ of data.
 We have carried out an independent analysis of the standard model backgrounds at $\sqrt s=7$ TeV and
 1~fb$^{-1}$ of integrated luminosity which are generally consistent with a previous study~\cite{Baer:2010tk}.
 {
 Our analysis is done  within the framework of the MSSM. However, the parameter space for soft breaking
 for the MSSM is rather large (with over a hundred parameters) and thus intractable.  This parameter  space
 is significantly reduced in the models we consider. Thus in the analysis of this work we use the framework  of
 mSUGRA which has four parameters and the sign of $\mu$, as well as  supergravity models  with
 nonuniversality in the gaugino sector which increases the number of parameters to six and the sign of
 $\mu$. The analysis is done under the imposition of theoretical constraints which include the
 radiative breaking of the electroweak symmetry, R parity conservation, and conservation of charge
 and color, as well as under the experimental constraints which include constraints on the relic density
 from WMAP, constraints from flavor changing neutral currents, and the experimental lower limit
 constraints on the masses of the  Higgs bosons and on the sparticle masses.
 The residual  model space which passes both the theoretical and the experimental constraints
 as described above is still large and we use a selection criteria in its further exploration.
 Noting that  the
  central theme of the analysis is the discovery of supersymmetry in early runs, our search for models
  is then narrowed to such models as are discoverable with $\sqrt s=7$ TeV and 1-2~fb$^{-1}$ of data.
  Within this general theme  we choose the models as broadly as possible
  to encompass as many diverse possibilities as possible.
 Specifically we select models so that all allowed NLSPs such as the chargino, the stau, the gluino,
 the CP odd Higgs, and the stop are included.
 Using the above criteria we have presented a set of benchmarks for early discovery in Table (III).
 We observe that in Table (III) all the NLSPs mentioned above are represented. Further, in  Table (III)  the range
 of inputs varies widely.  Thus, for example,
  $m_0$ ranges from $101$ to $2225$, $m_{1/2}$ ranges from $313$ to $755$, $A_0$
  ranges from $-2531$ to $2710$ (all masses in GeV), and $\tan\beta$ ranges from 5.7 to 47.2.
 The analysis of this work shows
  that  an {$\mh$} mass up to 400 GeV  and  $m_0$ up to  2 TeV could be accessible for certain combination{s}  of soft parameters. Further, the precise  set of discovery
 modes in which  the benchmarks  will become visible are identified.}
 It is shown that
 most of the benchmarks have at least a minimum of five discovery channels and  several of them have as many as 10
   in which the SUSY signal  is discoverable.
  We have also exhibited the  dependence of  the number of discovery channels as a function of the integrated
   luminosity which shows its rapid  increase as the integrated luminosity increases toward {1~fb$^{-1}$} for
   the set of discoverable models investigated.
  Thus  if SUSY is found in one of the signature channels  for any of the  benchmarks given in Table~(\ref{bench2}),
then  it should also show up in other signature channels as identified in Fig.~(\ref{grid})
 providing important cross-checks for the discovery of supersymmetry.
 All the benchmarks exhibited are consistent with the current limits on the spin independent neutralino-proton cross  section
 from XENON100 and CDMS-II dark matter detectors.  Further, all of the benchmarks will be accessible in the next generation of
 dark matter experiments.
   \\

\noindent
{\em Acknowledgments}:
Communications with Johan Alwall {and Timothy Stelzer} regarding MadGraph and with {Andre} Lessa regarding the
analysis of~\cite{Baer:2010tk} are acknowledged.
This research is supported in part by NSF Grants  No. PHY-0653587 and
No. PHY-0757959.

\section{Tables\label{tables}}

\begin{table}[h!]
\centering
{\bf \boldmath A display of signatures/cuts used in early discovery analysis at the LHC at $\sqrt s= 7$ TeV}
\vspace{.3cm}
\begin{tabular}{|c|c|c|c|}
\hline
& Signature Name & \multicolumn{2}{|c|}{Description of Cut}\\
\hline
 1 & monojets & $n(\ell)=0$  & $p_T(j_1) \geq 100 \GeV$, $p_T(j_2)<20 \GeV$  \\
2 & multi-jets200 & $n(\ell)=0$  & $p_T(j_1) \geq 200 \GeV$, $p_T(j_2) \geq 150 \GeV$, $p_T(j_4) \geq 50 \GeV$ \\
 3 & multi-jets100 & $n(\ell)=0$  & $p_T(j_1) \geq 100 \GeV$, $p_T(j_2) \geq 80 \GeV$, $p_T(j_4) \geq 40 \GeV$ \\
4 & hard-jets500 & $n(\ell)=0$  & $p_T(j_2) \geq 500 \GeV$ \\
5 & hard-jets350 & $n(\ell)=0$  & $p_T(j_2) \geq 350 \GeV$ \\
6 & multi-bjets1 & $n(\ell)=0$, $n(b) \geq 1$ &   \\
7 & multi-bjets2 & $n(\ell)=0$, $n(b) \geq 2$ &   \\
 8 & multi-bjets3 & $n(\ell)=0$, $n(b) \geq 3$ &   \\
 9 &$H_T500$ & $n(\ell)+n(j) \geq 4$  & $p_T(1) \geq 100 \GeV$ , $\sum_{i=1}^4 p_T(i) + \slashed{E}_T \geq 500 \GeV$  \\
10 &  $H_T400$ & $n(\ell)+n(j) \geq 4$  & $p_T(1) \geq 100 \GeV$ , $\sum_{i=1}^4 p_T(i) + \slashed{E}_T \geq 400 \GeV$  \\
11 & 1-lepton100 & $n(\ell)=1$  & $p_T(\ell_1) \geq 20 \GeV$, $p_T(j_1)\geq 100 \GeV$, $p_T(j_2)\geq 50 \GeV$  \\
12 & 1-lepton40 & $n(\ell)=1$  & $p_T(l_1) \geq 20 \GeV$, $p_T(j_2)\geq 40 \GeV$  \\
 13 & OS-dileptons100 & $n(\ell^+)=n(\ell^-)=1$  & $p_T(\ell_2) \geq 20 \GeV$, $p_T(j_1)\geq 100 \GeV$, $p_T(j_2)\geq 50 \GeV$  \\
14 & OS-dileptons40 & $n(\ell^+)=n(\ell^-)=1$  & $p_T(\ell_2) \geq 20 \GeV$, $p_T(j_2)\geq 40 \GeV$  \\
 15 & SS-dileptons100 & $n(\ell^+ \, | \, \ell^-)=n(\ell)=2$  & $p_T(\ell_2) \geq 20 \GeV$, $p_T(j_1)\geq 100 \GeV$, $p_T(j_2)\geq 50 \GeV$  \\
16 & SS-dileptons40 & $n(\ell^+ \, | \, \ell^-)=n(\ell)=2$  & $p_T(\ell_2) \geq 20 \GeV$, $p_T(j_2)\geq 40 \GeV$  \\
17 & 3-leptons100 & $n(\ell)=3$  & $p_T(l_3) \geq 20 \GeV$, $p_T(j_1)\geq 100 \GeV$, $p_T(j_2)\geq 50 \GeV$  \\
18 & 3-leptons40 & $n(\ell)=3$  & $p_T(l_3) \geq 20 \GeV$, $p_T(j_2)\geq 40 \GeV$  \\
19 & 4$^+$-leptons & $n(\ell)\geq 4$  & $p_T(l_4) \geq 20 \GeV$, $p_T(j_2)\geq 40 \GeV$  \\
20 & 1-tau100 & $n(\tau)=1$  & $p_T(\tau_1) \geq 20 \GeV$, $p_T(j_1)\geq 100 \GeV$, $p_T(j_2)\geq 50 \GeV$  \\
21 & 1-tau$40$ & $n(\tau)=1$  & $p_T(\tau_1) \geq 20 \GeV$, $p_T(j_2)\geq 40 \GeV$  \\
22 & OS-ditaus100 & $n(\tau^+)=n(\tau^-)=1$  & $p_T(\tau_2) \geq 20 \GeV$, $p_T(j_1)\geq 100 \GeV$, $p_T(j_2)\geq 50 \GeV$  \\
23 & OS-ditaus$40$ & $n(\tau^+)=n(\tau^-)=1$  & $p_T(\tau_2) \geq 20 \GeV$, $p_T(j_2)\geq 40 \GeV$  \\
24 & SS-ditaus100 & $n(\tau^+ \, | \, \tau^-)=n(\tau)=2$  & $p_T(\tau_2) \geq 20 \GeV$, $p_T(j_1)\geq 100 \GeV$, $p_T(j_2)\geq 50 \GeV$  \\
25 & SS-ditaus$40$ & $n(\tau^+ \, | \, \tau^-)=n(\tau)=2$  & $p_T(\tau_2) \geq 20 \GeV$, $p_T(j_2)\geq 40 \GeV$  \\
26 & 3$^+$-taus100 & $n(\tau) \geq 3$  & $p_T(\tau_3) \geq 20 \GeV$,  $p_T(j_1)\geq 100 \GeV$,             $p_T(j_2)\geq 50 \GeV$  \\
27 & 3$^+$-taus40 & $n(\tau)\geq 3$  & $p_T(\tau_4) \geq 20 \GeV$, $p_T(j_2)\geq 40 \GeV$  \\
28 & 1$^+$-photon & $n(\gamma)\geq 1$  & $p_T(j_2)\geq 40 \GeV$  \\
\hline
\end{tabular}
\caption{List of signatures and cuts used in the early discovery analysis. Our notation is as follows:
$\ell=e,\mu$, $n(x)$ is the number of object $x$ in the event, {and} $p_T(x_n)$ is the transverse momentum of the $n^{\text{th}}$
hardest object $x$. {For the case of $p_T(\tau)$ we take this to mean the visible part of the $p_T$ from a hadronically decaying tau.  }We required $\slashed{E}_T \geq 200 \GeV$ and a minimum transverse sphericity of 0.2. {The symbol $\mid$ should be read as the logic ``or'': i.e. the cut $n\left( \tau^{+}\mid\tau^{-}\right)=2$  would be read ``the number of $\tau^{+}$ equals 2 or the number of $\tau^{-}$ equals 2."
\label{SUSYcuts}
}}
\end{table}

\begin{table}[h!]

\centering
{\bf\boldmath Benchmarks for Early Discovery  at $\sqrt s=7$ TeV with 2~fb$^{-1}$ }\\
\vspace{.3cm}
\begin{tabular}{|x{1cm}|x{1cm}|x{1cm}x{1cm}x{1cm}x{1cm}x{1.2cm}x{1.2cm}|x{1.2cm}|c|} \hline
\multirow{2}{*}{Label} & \multirow{2}{*}{NLSP} & \multirow{2}{*}{$m_0$} & \multirow{2}{*}{$\mh$} & \multirow{2}{*}{$A_0$} &\multirow{2}{*}{$\tan\beta$} & \multirow{2}{*}{$\delta_2$ }& \multirow{2}{*}{$\delta_3$ } & $\sigma_{\rm SUSY}$ &  $\sigma_{\rm SI}$\tn
&  & & & & & &  &(pb) &$\lp 10^{-44}\text{cm}^2\rp$  \tnhl
\hline
C1 & $\chi^{\pm}_1$ & 1663 & 309 & 1508 & 32.9 & 0.553 & -0.687  &24.3& 7.0 \tn
C2 & $\chi^{\pm}_1$ & 449 & 330 & 176 & 20.3 & -0.382 & -0.151    & 2.4 & 3.7 \tn
C3 & $\chi^{\pm}_1$ & 1461 & 361 & 1327 & 30.3 & -0.241 & -0.702  &14.8 & 4.5 \tn
C4 & $ \chi^{\pm}_1$ & 1264 & 445 & 1775 & 24.7 & 0.718 & -0.736  &11.3 & 4.7 \tn
C5 & $ \chi^{\pm}_1$ & 240 & 313 & -522 & 5.48 & -0.376 & -0.106  &3.5& 0.7 \tn
G1 & $\wt{g}$ & 1694 & 755 & -2128 & 45.7 & 0.745 & -0.803 &2.2 & 4.9 \tn
 G2 & $\wt{g}$ & 2231 & 639 & 2710 & 18.0 & 0.543 & -0.850  & 24.2 & 3.0\tn
 G3 & $\wt{g}$ & 2276 & 615 & -2407 & 47.2& 0.631 & -0.784  & 3.1 & 2.6 \tn
G4& $\wt{g}$ & 2180 & 651 & -2271 & 47.1 & 0.680 & -0.817    &5.8 & 8.3\tn
G5 & $\wt{g}$ & 2126 & 683 & 2924 & 38.0 & 0.580 & -0.849 &19.4 & 4.8 \tn
G6 & $\wt{g}$ & 1983 & 749 & -2332 & 46.3 & 0.562 & -0.824 &3.7 & 2.7\tn
 H1 & $A^o$ & 2225 & 674 & -2531 & 47.3 & 0.783 & -0.703 &0.3 &0.9\tn
 S1 & $\wt{\tau}_1$ & 117 & 394 & 0 & 15.9 & -0.327 & -0.177&1.4 & 1.4 \tn
 S2 & $\wt{\tau}_1$ & 101 & 446 & -153 & 6.1 & 0.607 & -0.207& 0.4 &0.5\tn
 S3 & $\wt{\tau}_1$ & 102 &470& 183& 15.3 & 0.603 & -0.266 &0.5 &  3.0\tn
S4 & $\wt{\tau}_1$ & 309 & 581 & -613 & 27.7 & 0.839 & -0.400   & 0.6 & 1.6\tn
S5 & $\wt{\tau}_1$ & 135 & 688 & -184 &5.7 & -0.052 &-0.499   & 0.4 & 1.6\tn
S6 & $\wt{\tau}_1$ & 114 & 404 & 27 & 13.0 & -0.369 & -0.267  & 2.0 & 3.0 \tn
S7 & $\wt{\tau}_1$ & 114 & 518 & 87& 10.4 & 0.266 & -0.247 &  0.2 &0.6\tn
T1 & $\wt{t}_1$ & 1726 & 548 & 4197 & 21.2 & 0.132 & -0.645 &  2.3 &5.0$\times 10^{-3}$\tn
 T2 & $\wt{t}_1$& 1590 & 755 & 3477 & 23.4 & 0.805 & -0.803 &  3.8&9.4$\times 10^{-2}$\tnhl
 \end{tabular}\caption{ {Benchmarks for {\it    models discoverable} at the LHC at $\sqrt s=7$ TeV
with 2~fb$^{-1}$ of integrated luminosity. The model
 inputs are given  at $M_{GUT}=2\times10^{16}\text{ GeV}$,
  sign$(\mu)=+1$, {and} $\delta_1=0$. The displayed  masses are in GeV.  All models satisfy
  REWSB and the experimental constraints as discussed in Sec.~(\ref{background}).  {The spin independent direct detection cross section, $\sigma_{\text{SI}}$, is exhibited as well as the
  cross section $\sigma_{\text{SUSY}}$ for the production of supersymmetric particles
    at $\sqrt{s}=7$ TeV.}
  Our analysis shows that all the models listed in this table
  are discoverable at the 5$\sigma$ level above  the
background in  {\it several channels}  as exhibited in Fig.~(\ref{grid}).
   }
\label{bench2}}
\end{table}

\begin{table}[h!]
\centering
{\bf Sparticle mass spectra for the benchmarks}
\vspace{0.5cm}
\begin{tabular}{|l|x{1cm}|x{1cm}|x{1cm}|x{1cm}|x{1cm}|x{1cm}|x{1cm}|x{1cm}|x{1cm}|x{1cm}|x{1cm}|x{1cm}|}
\hline
Label &   C1 & C2 & C3  & C4  &   C5 & G1& G2 &  G3 &G4 &G5&G6 \tnhl
NLSP& $\chi^{\pm}_1$ &$\chi^{\pm}_1$ &$\chi^{\pm}_1$ & $ \chi^{\pm}_1$ & $ \chi^{\pm}_1$ &$\tilde{g}$&$\tilde{g}$&$\tilde{g}$&$\tilde{g}$&$\tilde{g}$&$\tilde{g}$\tnhl
$\sigma_{\text{SUSY}}$ (pb)& 24.3&2.4&14.8  & 11.3& 3.5& 2.2 &24.2 & 3.1& 5.8&19.4&3.7\tn
 $\mu$& 145& 345& 239&231 & 489 &480 &314 & 523&434&324&539\tn
 $m_{\tilde{N}_1}$& 103& 130& 140 & 171 & 123&327 &256& 270&283&275&328 \tn
 $m_{\tilde{N}_2}$& 157 &151& 189& 240 & 146& 485&324 & 522&439&333&541\tn
 $m_{\tilde{C}^{\pm}_1}$& 141& 150&183 & 229 & 145& 480&316 & 520&435&326&539 \tn
 $m_{{\stau}_1}$& 1463& 441 &1318& 1170 &260& 1079 &2155 & 1542&1475&1719&1311 \tn
 $m_{\tilde{t}_1}$& 922&560&828 & 621 & 402&612 &1145 &  1115&1048 & 1054&863\tn
 $m_{\tilde{g}}$&316&698&341& 354 &680& 452&314  & 432&393&325 &421\tnhl
\end{tabular}

\centering
 \begin{tabular}{|l|x{1cm}|x{1cm}|x{1cm}|x{1cm}|x{1cm}|x{1cm}|x{1cm}|x{1cm}|x{1cm}|x{1cm}|x{1cm}|x{1cm}|}
\hline
 Label &  H1& S1 & S2 & S3&S4&S5&  S6&   S7     &  T1 & T2\tnhl
 NLSP &   $A^o$& $\tilde{\tau}_1$&$\tilde{\tau}_1$ &$\tilde{\tau}_1$&$\tilde{\tau}_1$&$\tilde{\tau}_1$& $\tilde{\tau}_1$&$\tilde{\tau}_1$     &$\tilde{t}_1$&$\tilde{t}_1$\tnhl
 $\sigma_{\text{SUSY}}$ (pb)&    0.3&1.4&0.4& 0.5&0.6& 0.4&2.0& 0.2 &2.3& 3.8\tn
$\mu$&   630& 425& 426& 301& 393&443&  388& 432&1213&691\tn
 $m_{\tilde{N}_1}$& 297&157& 181 & 188&241&280& 159& 211&232&324 \tn
 $m_{\tilde{N}_2}$& 630&198&408&  303&395& 415& 189& 408 &508&689 \tn
$m_{\tilde{C}_1^{\pm}}$& 629&198&405& 295&390&409& 187& 406 &508&687\tn
$m_{{\stau}_1}$& 1456&167&194& 192&248&289& 176& 221 &1546&1436 \tn
$m_{\tilde{t}_1}$&  1032&541&504& 529&329&506& 497& 615&258&357\tn
$m_{\tilde{g}}$& 594&771&835& 817&834&817 &709& 913&532&422\tnhl
\end{tabular}

\caption{An exhibition of  the light sparticles for the benchmarks given.  These benchmarks are listed in Table~(\ref{bench2}).  {All masses are given in $\GeV$.}}
\label{susy}
\end{table}

\begin{table}[t!]
\centering
{\bf {LHC discovery channels after 1~fb$^{-1}$ of integrated luminosity}}
\vspace{.25cm}
\begin{center}
\begin{tabular}{|l|x{1.7cm}||x{1.3cm}||x{1.3cm}|x{1.3cm}|x{1.3cm}|x{1.3cm}|x{1.3cm}|x{1.3cm}|x{1.3cm}|x{1.3cm}|}
\hline
Signature Name~~~& &  SM & C1 & C4 & C5 & G2 & G3 & S3 & S6 & T1
\tnhl
%
\multirow{2}{*}{Multi-jets200}& Events & 47 & 91 & 68 & 105 & 28 & 16 & 49 & 88 & 12 \tn
&  $S/\sqrt{B}$ & $\cdots$ & 13.3 & 9.9 & 15.2 & 4.1 & 2.4 & 7.2 & 12.7 &
1.8 \tnhl
\multirow{2}{*}{Multi-jets100}&  Events & 180 & 401 & 225 & 213 & 114 & 83 & 77 & 171 & 69 \tn
 & $S/\sqrt{B}$ & $\cdots$ & 29.9 & 16.8 & 15.9 & 8.5 & 6.2 & 5.8 & 12.8 &
5.2 \tnhl
\multirow{2}{*}{Multi-jets40}  & Events & 215 & 497 & 316 & 218 & 135 & 107 & 77 & 176 & 85 \tn
 & $S/\sqrt{B}$ & $\cdots$ & 33.9 & 21.6 & 14.9 & 9.2 & 7.3 & 5.3 & 12.0 &
5.8 \tnhl
\multirow{2}{*}{$H_T$400} &Events & 965 & 1035 & 501 & 496 & 286 & 183 & 156 & 419 & 143 \tn
 & $S/\sqrt{B}$ & $\cdots$ & 33.3 & 16.1 & 16.0 & 9.2 & 5.9 & 5.0 & 13.5 &
4.6 \tnhl
\multirow{2}{*}{Multi-bjets1}& Events & 188 & 460 & 188 & 175 & 51 & 126 & 50 & 102 & 86 \tn
 & $S/\sqrt{B}$ & $\cdots$ & 33.5 & 13.7 & 12.8 & 3.7 & 9.2 & 3.6 & 7.5 &
6.3 \tnhl
\multirow{2}{*}{Multi-bjets2} & Events & 46 & 157 & 49 & 69 & 7 & 57 & 19 & 39 & 45 \tn
 & $S/\sqrt{B}$ & $\cdots$ & 23.1 & 7.3 & 10.1 & $\cdots$ & 8.4 & 2.8 & 5.7 & 6.6
\tnhl
\multirow{2}{*}{1-lepton40}&  Events & 367 & 45 & 20 & 74 & 0 & 0 & 30 & 38 & 27 \tn
 & $S/\sqrt{B}$ & $\cdots$ & 2.4 & 1.0 & 3.9 & $\cdots$ & $\cdots$ & 1.6 & 2.0 & 1.4
\tnhl
\end{tabular}
{\caption{\label{discovery}  LHC discovery channels after 1~fb$^{-1}$
of integrated luminosity for selected benchmark models from
Table~(\ref{susy}). All signatures require transverse sphericity
$S_T \geq 0.2$ and at least 200~GeV of $\slashed{E}_T$. For each
signature the number of signal events is given, as well as the
signal significance if there are sufficient signal events. We have
also included a much weaker multijet signature (multijets40) in
which the four jets are all required merely to satisfy $p_T^{\rm
jet} \geq 40\GeV$. This signature also appears in
Fig.~(\ref{graphcuts}).}}
\end{center}
\end{table}

\end{document}